\documentclass[useAMS,usenatbib]{mn2e}
\usepackage{amsmath,amssymb}
\usepackage{graphicx,subfigure}
\usepackage{multirow}
\pdfoutput=1

\newcommand{\bb}{\begin{bmatrix}}
\newcommand{\eb}{\end{bmatrix}}

\title[Searching for continuous GWs in PPTA data]{An all-sky search for continuous gravitational waves in the Parkes Pulsar Timing Array data set}

\author[X.-J. Zhu et al.]{X.-J. Zhu$^{1,2}$\thanks{E-mail: zhuxingjiang@gmail.com},
 G. Hobbs$^{2}$,
 L. Wen$^{1}$,
 W. A. Coles$^{3}$,
 J.-B. Wang$^{4,2}$,
 R. M. Shannon$^{2}$,
 \newauthor
 R. N. Manchester$^{2}$,
 M. Bailes$^{5}$,
 N. D. R. Bhat$^{6}$,
 S. Burke-Spolaor$^{7}$,
 S. Dai$^{8,2}$,
 \newauthor
 M. J. Keith$^{9}$,
 M. Kerr$^{2}$,
 Y. Levin$^{10}$,
 D. R. Madison$^{11}$,
 S. Os{\l}owski$^{12,13}$,
 \newauthor
 V. Ravi$^{14,2}$,
 L. Toomey$^{2}$,
 W. van Straten$^{5}$\\
$^{1}$School of Physics, University of Western Australia, Crawley WA 6009, Australia\\
$^2$ CSIRO Astronomy and Space Science, PO Box 76, Epping NSW 1710, Australia\\
$^3$ Department of Electrical and Computer Engineering, University of California at San Diego, La Jolla, CA 92093, USA\\
$^4$ Xinjiang Astronomical Observatory, CAS, 150 Science 1-Street, Urumqi, Xinjiang 830011, China\\
$^5$ Centre for Astrophysics and Supercomputing, Swinburne University of Technology, PO Box 218, Hawthorn Vic 3122, Australia\\
$^6$ International Centre for Radio Astronomy Research, Curtin University, Bentley, WA 6102, Australia\\
$^7$ Jet Propulsion Laboratory, California Institute of Technology, 4800 Oak Grove Dr, Pasadena CA 91109-8099, USA\\
$^8$ Department of Astronomy, School of Physics, Peking University, Beijing, 100871, China\\
$^9$ Jodrell Bank Centre for Astrophysics, School of Physics and Astronomy, University of Manchester, Manchester M13 9PL, UK\\
$^{10}$ School of Physics, Monash University, PO Box 27, Vic 3800, Australia\\
$^{11}$ Department of Astronomy and Center for Radiophysics and Space Research, Cornell University, Ithaca, NY 14850, USA\\
$^{12}$ Max-Planck-Institut f{\"u}r Radioastronomie, Auf dem H{\"u}gel 69, D-53121 Bonn, Germany\\
$^{13}$ Department of Physics, Universit{\"a}t Bielefeld, Universit{\"a}tsstr. 25 D-33615 Bielefeld, Germany\\
$^{14}$ School of Physics, University of Melbourne, Vic 3010, Australia\\}

\begin{document}
\date{}
\pagerange{\pageref{firstpage}--\pageref{lastpage}} \pubyear{2014}
\maketitle \label{firstpage}

\begin{abstract}
We present results of an all-sky search in the Parkes Pulsar Timing Array (PPTA) Data Release 1 data set for continuous gravitational waves (GWs) in the frequency range from $5\times 10^{-9}$ to $2\times 10^{-7}$ Hz. Such signals could be produced by individual supermassive binary black hole systems in the early stage of coalescence. We phase up the pulsar timing array data set to form, for each position on the sky, two data streams that correspond to the two GW polarizations and then carry out an optimal search for GW signals on these data streams. Since no statistically significant GWs were detected, we place upper limits on the intrinsic GW strain amplitude $h_0$ for a range of GW frequencies. For example, at $10^{-8}$ Hz our analysis has excluded with $95\%$ confidence the presence of signals with $h_0\geqslant 1.7\times 10^{-14}$. Our new limits are about a factor of four more stringent than those of Yardley et al. (2010) based on an earlier PPTA data set and a factor of two better than those reported in the recent Arzoumanian et al. (2014) paper. We also present PPTA directional sensitivity curves and find that for the most sensitive region on the sky, the current data set is sensitive to GWs from circular supermassive binary black holes with chirp masses of $10^{9} M_{\odot}$ out to a luminosity distance of about 100 Mpc. Finally, we set an upper limit of $4 \times 10^{-3} {\rm{Mpc}}^{-3} {\rm{Gyr}}^{-1}$ at $95\%$ confidence on the coalescence rate of nearby ($z \lesssim 0.1$) supermassive binary black holes in circular orbits with chirp masses of $10^{10}M_{\odot}$.
\end{abstract}
\begin{keywords}
gravitational waves -- methods: data analysis -- pulsars: general -- galaxies: evolution
\end{keywords}

\section{INTRODUCTION}
Pulsar timing arrays (PTAs) have long been proposed to detect very low frequency ($10^{-9}$ to $10^{-7}$ Hz) gravitational waves \citep[GWs;][]{Hellings_Downs83,Foster_Backer90}. The concept of a PTA is to conduct long-term timing observations of a number of spatially separated millisecond pulsars whose pulsational periods are extremely stable. GWs sweeping over the pulsar or the Earth introduce fluctuations in pulse times of arrival (ToAs) -- named ``pulsar terms" and ``Earth terms" respectively. By comparing the measured ToAs to predictions from a timing model that accounts for the pulsar's intrinsic rotation and radio pulse propagation \citep{TEMPO2Edwards}, timing residuals are formed and can be searched for GWs. Currently three PTAs are in operation: the Parkes PTA \citep[PPTA;][]{PPTA2013,PPTA13CQG}, the European PTA \citep[][]{EPTA}, and NANOGrav \citep{NANOGrav}, each of which has collected timing data for $\sim 20$ ms pulsars with precision from $\mu$s down to tens of ns and data spans of $\gtrsim 5$ yrs. It is expected that these PTA data sets will be combined to form the initial International PTA \citep{IPTA,IPTAdick13} data set in the near future.

One of the most promising source classes in the PTA frequency range is inspiralling super-massive binary black holes (SMBBHs) in the centres of galaxies. Previous studies, including both theoretical modelling \citep[e.g.][]{Jaffe_Backer03,Wyithe_Loeb03,Sesana13GWB,Ravi14GWB} and actual analyses of real PTA data \citep{Jenet2006,YardleySGWB,EPTAlimit,NANOGrav2012,PPTA2013Sci}, have mostly focused on the stochastic background formed by the superposition of emission from a large number of single sources. However, individual resolvable sources that are sufficiently close and/or massive may provide chances for the detection of continuous waves \citep[CWs;][]{Sesana2009,Ravi2012,Ravi14Single}.

There has recently been a lot of work on PTA data analysis methods for CW detection and parameter estimation \citep{Yardley2010,CorbinCornish10,KJLee2011,Babak2012,Ellis2012,EllisBayesian,Petiteau13,NANOGravCW14,Taylor14,YanWang14}, ranging from frequentist to Bayesian techniques and from Earth-term-only approaches to a coherent inclusion of both Earth terms and pulsar terms. In particular, \citet[][PPTA10 hereafter]{Yardley2010} used an earlier PPTA data set presented in \citet{Verbiest09} to produce sensitivity curves and set upper limits with a power spectral summation method, and almost simultaneously to this work \citet{NANOGravCW14} applied both frequentist \citep{Ellis2012} and Bayesian \citep{EllisBayesian} data analysis pipelines to the NANOGrav 5-year, 17 pulsar data set \citep{NANOGrav2012} to compute upper limits on the strain amplitudes of CWs from circular SMBBHs.

In this paper we have developed a new method which conducts a global fit for Earth-term timing residuals induced by single-source GWs (regardless of their specific waveforms) and outputs two time series which correspond to the two GW polarizations and their covariance matrix. A maximum likelihood detection technique is then applied to the two time series to form our detection statistics for CWs from different sky directions. Using this method we perform an all-sky search for CWs that could be produced by circular SMBBHs in the PPTA Data Release 1 (DR1) data set \citep{PPTA2013}. As the search did not find any statistically significant CW events, we set upper limits on the intrinsic GW strain amplitude $h_0$ and present all-sky and directional sensitivity curves.

The organization of this paper is as follows. In section \ref{sec:obser} we provide a brief overview of our timing observations. In section \ref{sec:metho} we describe the signal model and analysis methods used in performing the search. In section \ref{sec:results} we present our results and discussions. Finally we conclude in section \ref{sec:conclu}.
\vspace{-5mm}
\section{OBSERVATIONS}
\label{sec:obser}
Based on the Parkes 64-m radio telescope, the PPTA project started in early 2005 routine timing observations (e.g., once every 2-3 weeks) of 20 ms pulsars in three radio bands (10, 20 and 50 cm) with a typical integration time of 1 h. With the development of new instrumentation, the timing precision has been steadily improved \citep[see][for details]{PPTA2013}. We use in this search the DR1 data set reported\footnote{A slight modification was made to the published DR1 data set to fix a small offset in early 10 cm (3 GHz) data for J1909$-$3744 \citep{PPTA2013Sci}.} in \citet{PPTA2013}, including observations made between 2005 March 1 (MJD 53430) and 2011 February 28 (MJD 55620). For each pulsar ToAs of the best band (i.e., where the lowest rms timing residuals are seen) have been selected after correcting for dispersion measure (DM) variations \citep{MikeDM2013}. The PPTA DR1 data set is publicly available online at a permanent link\footnote{\url{http://dx.doi.org/10.4225/08/534CC21379C12}}.

The data set that we use here is identical to that used for searching for GW memory events by \citet{GWM_PPTA}. Details on how we obtain the noise models for each pulsar are given in that paper and we only give a very brief description here. For about half the PPTA pulsars, low-frequency timing noise (``red noise") is observed in the timing residuals. In the case of detected red noise, we first fit a power-law model to the power spectrum of timing residuals, then obtain estimates of noise covariance matrices iteratively, and finally use the Cholesky decomposition of this covariance matrix to transform the problem to an ordinary least-squares problem \citep{Coles2011}. Additionally, it is common that white noise of the timing residuals is underestimated by their ToA uncertainties. We therefore introduce the factors -- ``EFAC" and ``EQUAD" in \textsc{TEMPO2} \citep{TEMPO2} -- to rescale the ToA errors so that the observed scatter is represented. Since our analysis depends on accurate noise models, we have used simulations to show that the mean power spectra obtained from data sets simulated with those noise models agree with the power spectra obtained from the actual data sets. These simulations are carried out with the same data span and sampling as in the actual PPTA DR1 data. Throughout our analysis we use a ``fixed-noise" approach which means that 1) noise models are determined before the CW search is implemented and 2) we do not repeat the noise estimation process in the case of signal injections.
\vspace{-5mm}
\section{THE DATA ANALYSIS METHOD}
\label{sec:metho}
\subsection{The signal model}
\label{Sig_model}
The ToA variations induced by a single-source GW can be generally written as:
\begin{equation}
\label{TR1}
r(t,\hat{\Omega}) = F_{+}(\hat{\Omega})\Delta A_{+}(t) + F_{\times}(\hat{\Omega}) \Delta A_{\times}(t),
\end{equation}
where $\hat{\Omega}$ is a unit vector defining the direction of GW propagation. The two functions $F_{+}(\hat{\Omega})$ and $F_{\times}(\hat{\Omega})$ are the geometric factors, equivalent to the antenna pattern functions used in the context of laser interferometric GW detection \citep{Thorne87}, which only depend on the GW source position for a given pulsar as given by \citep{KJLee2011}:
\begin{eqnarray}
F_{+}(\hat{\Omega}) &=& \frac{1}{4(1-\cos\theta)}\{(1+\sin^2 \delta)\cos^2 \delta_p \cos[2(\alpha-\alpha_p)]\nonumber\\&&\hspace{-14mm} - \sin2\delta \sin2\delta_p\cos(\alpha-\alpha_p) + \cos^2 \delta (2-3\cos^2 \delta_p)\}
\label{Fp}
\end{eqnarray}
\begin{eqnarray}
F_{\times}(\hat{\Omega}) &=& \frac{1}{2(1-\cos\theta)}\{\cos \delta \sin 2\delta_p \sin(\alpha-\alpha_p)\nonumber\\&& - \sin \delta \cos^2 \delta_p \sin[2(\alpha-\alpha_p)]\} ,
\label{Fc}
\end{eqnarray}
where $\cos\theta = \cos\delta \cos\delta_p \cos(\alpha-\alpha_p)+\sin\delta \sin\delta_p$ with $\theta$ being the angle between the GW source direction and pulsar direction with respect to the observer, $\delta$ ($\delta_p$) and $\alpha$ ($\alpha_p$) are the declination and right ascension of the GW source (pulsar) respectively.

As the GW induced pulsar ToA variations result from an integration effect of the metric perturbation along the path from the pulsar to the Earth, they can be expressed as the combination of two terms -- the Earth term $A_{+,\times}(t)$ and the pulsar term $A_{+,\times}(t_p)$ \citep[see, e.g.,][]{Hellings81,JenetWen04}:
\begin{equation}
\label{TR2}
\Delta A_{+,\times}(t) = A_{+,\times}(t)-A_{+,\times}(t_p)
\end{equation}
\begin{equation}
\label{TpTe}
t_p = t-d_p(1-\cos\theta)/c,
\end{equation}
where $d_p$ is the pulsar distance and we have adopted the plane wave approximation\footnote{\citet{DengXH2011} suggested that it is possible to measure distances to GW sources by considering corrections to this approximation for sources closer than $\sim$ 100 Mpc when sub-pc precision distance measurements to pulsars are assumed. Such corrections would have negligible impact on our results as we only search for Earth-term signals.}. In this work we focus on the correlated Earth-term signals and treat the incoherent pulsar terms as an extra source of noise, leaving the investigation on how pulsar terms could be incorporated to improve the detectability and angular resolution to a future study. $A_{+}(t)$ and $A_{\times}(t)$ are source-dependent functions, and at Newtonian order take the following forms for CWs emitted by SMBBHs in circular orbits \citep{Babak2012,Ellis2012}:
\begin{eqnarray}
A_{+}(t) &=& \frac{h_0}{2\pi f(t)} \{(1+\cos ^{2}\iota) \cos2\psi \sin[\Phi(t)+\Phi_0]\nonumber\\&& +2\cos\iota \sin2\psi \cos[\Phi(t)+\Phi_0]\}
\label{Ap}
\end{eqnarray}
\begin{eqnarray}
A_{\times}(t) &=& \frac{h_0}{2\pi f(t)} \{(1+\cos ^{2}\iota) \sin2\psi \sin[\Phi(t)+\Phi_0]\nonumber\\&& -2\cos\iota \cos2\psi \cos[\Phi(t)+\Phi_0]\},
\label{Ac}
\end{eqnarray}
where $\iota$ is the inclination angle of the binary orbit with respect to the line of sight, $\psi$ is the GW polarization angle, $\Phi_0$ is a phase constant, and the intrinsic GW strain amplitude $h_0$ is given by
\begin{equation}
h_{0}=2\frac{(G M_c)^{5/3}}{c^{4}}\frac{(\pi f)^{2/3}}{d_{L}}
\label{h0},
\end{equation}
where $d_L$ is the luminosity distance of the source, and the chirp mass $M_c$ is defined as $M_c^{5/3} = m_1 m_2(m_1+m_2)^{-1/3}$ with $m_1$ and $m_2$ being the binary component masses. It should be noted that we have neglected effects of redshift as the current data set is only sensitive to SMBBHs up to $z \sim 0.1$ even for the most massive sources. The GW phase and frequency are given by:
\begin{equation}
\label{GW-Phase}
\Phi(t) = \frac{1}{16}  \left(\frac{G M_c}{c^{3}}\right)^{-5/3}\left\{(\pi f_{0})^{-5/3}-[\pi f(t)]^{-5/3}\right\},
\end{equation}
\begin{equation}
\label{GW-freq}
f(t) = \left[f_{0}^{-8/3}-\frac{256}{5}\pi^{8/3}\left(\frac{G M_c}{c^{3}}\right)^{5/3}t\right]^{-3/8},
\end{equation}
where $f_0$ is the GW frequency at the time of our first observation. Unless we will later otherwise specify, we assume throughout the paper that 1) the source frequency evolution within a typical observation span of $\sim 10$ years is negligible. This is appropriate for most of the observable sources based on current astrophysical predictions \citep{Sesana2010}, therefore the above two equations are reduced to $\Phi(t) \simeq 2\pi f_{0} t$ and $f(t)\simeq f_{0}$ (will later simply use $f$ to denote GW frequency); 2) pulsar terms are in the same frequency bin as Earth terms\footnote{It is widely considered that these two terms are well separated in frequency for an astrophysically plausible sample of GW sources, as shown in fig. 2 in \citet{Sesana2010}. However, we note that the approximation in their equation 30 is not valid for pulsar-Earth light-travel time of Kyrs and thus overestimated the frequency separation. For example, for the typical parameter values the correct equation gives a separation of 9.83 nHz in contrast to 15 nHz given in their equation 30, while the frequency bin width in the current analysis is 1.32 nHz.}, as also assumed in PPTA10. Assumption 2) applies in the case of non-evolving sources. As our detection method targets only at the Earth-term signals, pulsar terms only matter in the establishment of upper limits by acting as a ``self-generated" source of noise. For completeness we also relax the assumption 2) and show that it does not affect our results significantly. We use the pulsar distance ($d_p$) estimates derived from DM in the ATNF Pulsar Catalogue\footnote{http://www.atnf.csiro.au/research/pulsar/psrcat/} \citep{ATNF05Pulsar} -- the actual choice of $d_p$ has negligible effects on our results.
\vspace{-5mm}
\subsection{Accounting for effects of single-source GWs in \textsc{TEMPO2}}
\label{fitApc}
As shown in the previous section, single-source GWs affect ToAs in a deterministic and quadrupolar way. For any source sky position, the coherent Earth-term GW signals $A_{+,\times}(t)$ can be considered as common signals existing in multiple pulsars. Using a method similar to \citet{George_Clock12} who searched for a common signal in all the pulsar data sets, we fit for $A_{+,\times}(t)$ as a set of equally spaced samples without specifying their functional forms but making use of the geometric factors given in equations (\ref{Fp}-\ref{Fc}). Linear interpolation is used in such a global fit as observations for different pulsars are usually unevenly sampled and not at identical times. To avoid introducing additional notation, hereafter we will refer to the two time series estimated with \textsc{TEMPO2} as $A_{+,\times}(t)$, which could contain potential CW signals of the form given by equations (\ref{Ap}-\ref{Ac}). An example of the $A_{+,\times}$ fit for burst GWs is given in fig. 5 of \citet{PPTA13CQG}. A complete discussion on this method and its applications to single-source GW detection will be detailed in a forthcoming paper (Madison et al., in preparation). Below we briefly discuss some features of this method. In the next section we outline a maximum likelihood technique for CW detection in $A_{+,\times}(t)$.

The $A_{+,\times}$ fit has the following properties: a) it allows one to simultaneously fit for single-source GWs and normal pulsar timing parameters; b) it significantly reduces the computation because the number of data points in $A_{+,\times}(t)$ is typically $\sim 100$ in comparison to thousands of ToAs for data spans of $\lesssim$ 10 yrs; c) it can be used to check the correctness of the noise models. For example, if our pulsar noise models are correct and the covariance matrix estimation is reliable, the whitened $A_{+,\times}(t)$ time series independently follow the standard Gaussian distribution in the absence of signals. The property a) also requires that constraints must be set on $A_{+,\times}(t)$ to avoid the covariance between a global fit for $A_{+,\times}(t)$ and the fit for timing model parameters of individual pulsars. Currently three kinds of constraints are implemented in \textsc{TEMPO2}, namely, 1) quadratic constraints that correspond to pulsar spin parameters, 2) annual sinusoids for pulsar positions and proper motions, 3) biannual sinusoids for pulsar parallax. These were first introduced and implemented in \citet{MikeDM2013} where the DM variations were modelled as linear interpolants. Details on the constrained least-squares fitting can be found in Appendix A of \citet{MikeDM2013}.

\begin{figure}
\begin{center}
        \subfigure[]{%
            \label{fig:ResCW}
            \includegraphics[width=0.46\textwidth]{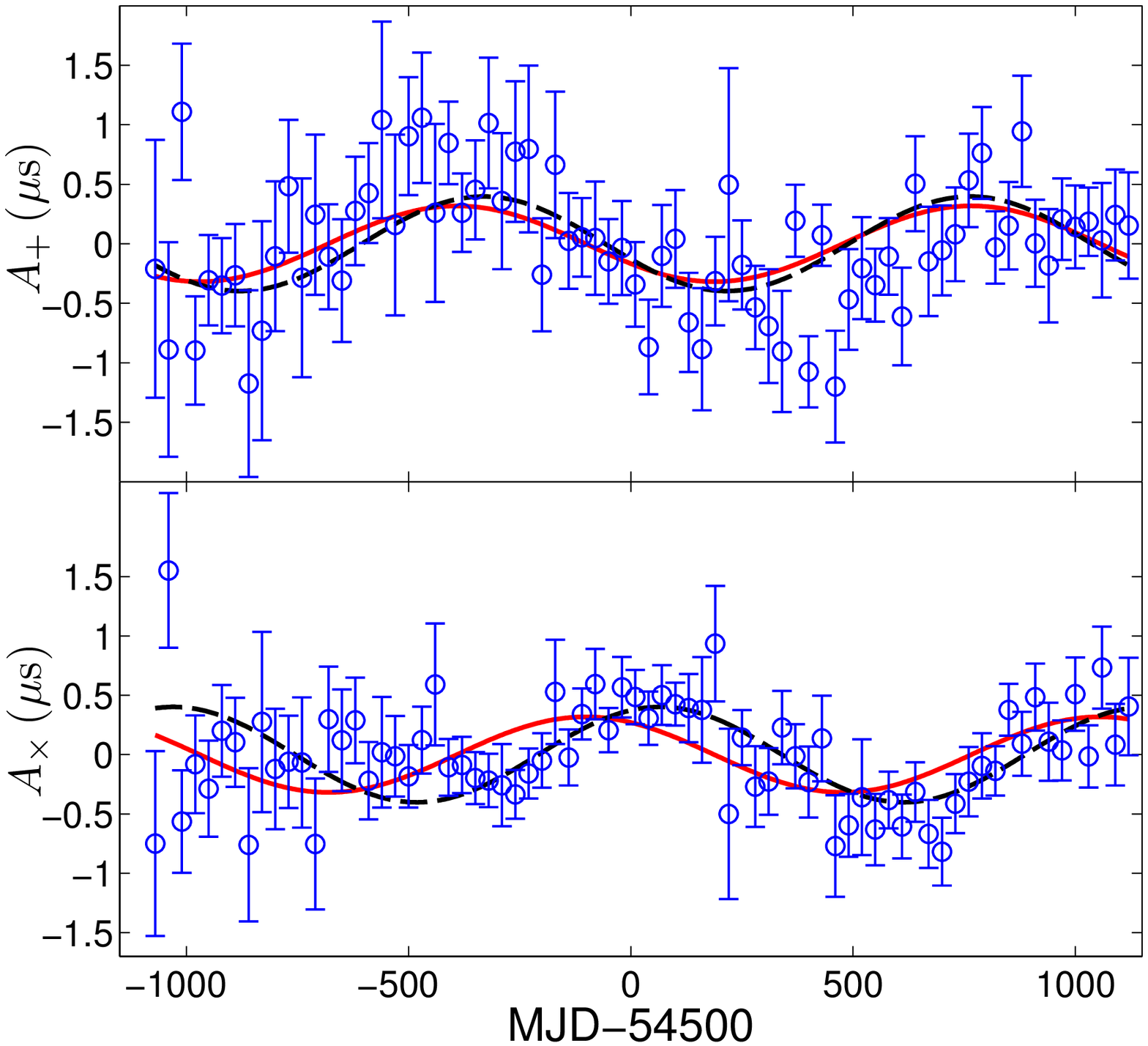}
        }\\
        \subfigure[]{%
            \label{fig:ApcFit}
            \includegraphics[width=0.46\textwidth]{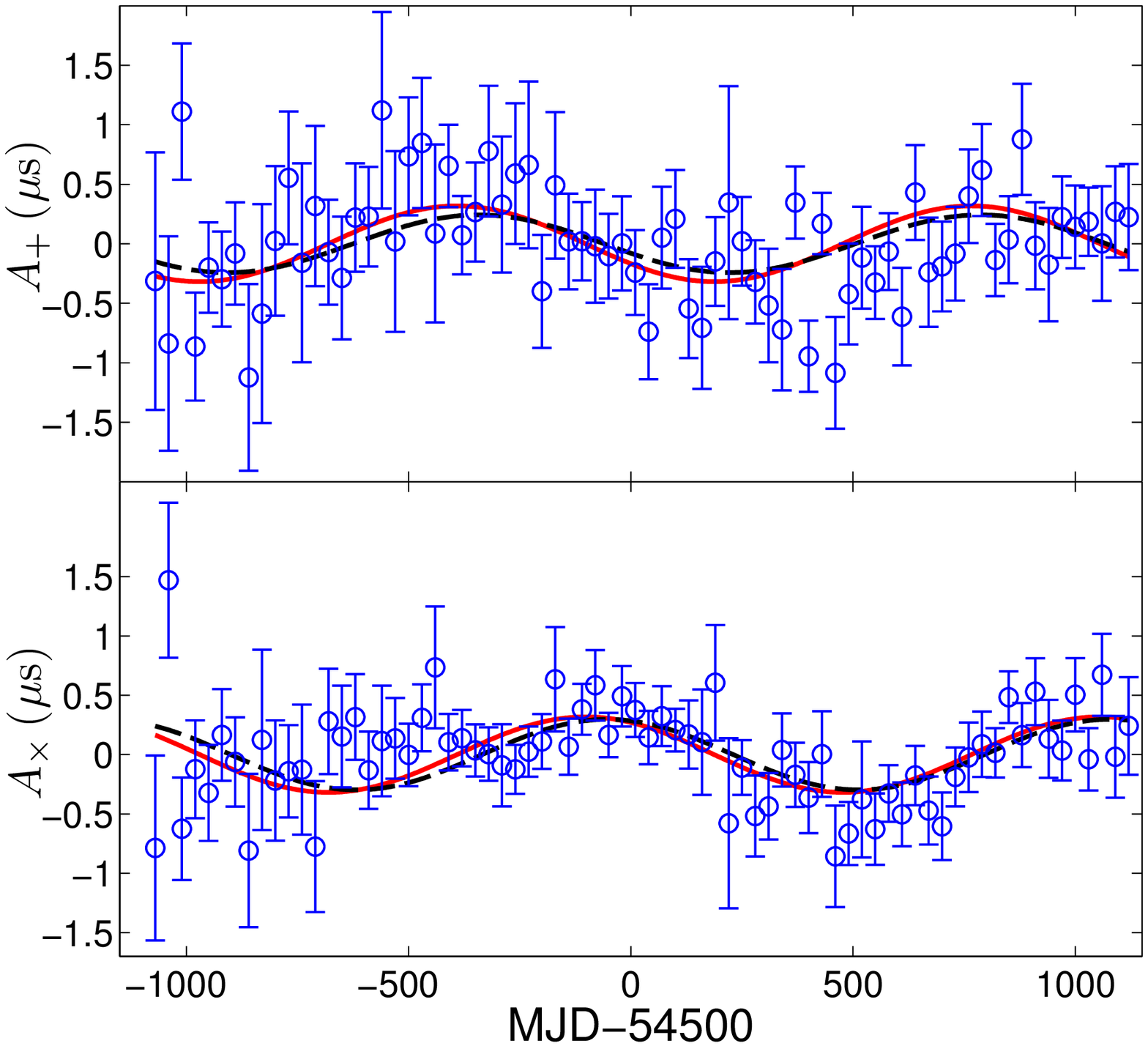}
        }
\end{center}
\caption{(a) The $A_{+,\times}(t)$ time series (open circles with error bars) estimated with a global least-squares fit in the PPTA DR1 data set that has included a CW signal injection (see text). The injected waveforms are plotted as solid red lines and the reconstructed waveforms based on a maximum likelihood technique are depicted by black dash lines. (b) As panel (a) but for the case of without pulsar terms in the signal injection.}
\label{fig:ApcCW}
\end{figure}

For the purpose of illustration of our method, we inject to the PPTA DR1 data set a CW signal\footnote{This functionality is available with the \textsc{TEMPO2} \textit{addCGW} plugin.} specified with the following parameters: $M_{c}=7.35\times 10^{8} M_{\odot}$, $d_{L}=16.5$ Mpc, $f=10$ nHz, $\cos\iota=1$, $\psi=\Phi_{0}=0$, $\alpha=3.2594$ and $\delta=0.2219$ (located in the Virgo cluster). Note that: 1) the chirp mass is chosen such that $h_{0}=10^{-14}$, which is below our upper limit ($1.7 \times 10^{-14}$) at 10 nHz; 2) the frequency separations between pulsar terms and Earth terms for different pulsars are determined physically using equations (\ref{TpTe}) and (\ref{GW-freq}). In this new data set we then fit for a full timing model for each pulsar and globally for $A_{+,\times}(t)$ at the injected sky location and for evenly spaced times between MJD 53430--55620 with a sampling interval of 30 d. The resulting estimates of $A_{+,\times}(t)$ are displayed as open circles with error bars in panel (a) of Fig. \ref{fig:ApcCW}, while the injected waveforms are shown as solid red lines. Because of the constraints applied in the global fit, data points in $A_{+,\times}(t)$ are known to be degenerated and thus it is necessary to perform a maximum likelihood estimation of two sinusoidal waves using the noise covariance matrix (e.g., as we outline in the next section). In this case, the frequency is estimated to be 10.6 nHz and the reconstructed waveforms are plotted as black dash lines. One can see the phase and amplitude are biased because of pulsar terms. To more clearly see this effect, we remove the pulsar terms in the signal injection and redo the same analysis. As shown in panel (b) of Fig. \ref{fig:ApcCW}, the match-up becomes much better. In the Earth-term-only case, the maximum likelihood frequency is 10.2 nHz. We leave the investigation of parameter estimation to a future study as in the current work we focus on detection and sensitivities.
\vspace{-5mm}
\subsection{Searching for CWs in $A_{+}(t)$ and $A_{\times}(t)$ time series}
\label{MLEapc}
Now the search for CWs is ready to be performed in $A_{+}(t)$ and $A_{\times}(t)$ time series. In general, these two time series data can be written as:
\begin{equation}
\mathbf{d}=\bb \mathbf{d}_{+} \\ \mathbf{d}_{\times} \eb=\bb \mathbf{s}_{+} \\ \mathbf{s}_{\times} \eb + \bb \mathbf{n}_{+} \\ \mathbf{n}_{\times} \eb,
\end{equation}
where $\mathbf{s_{+,\times}}$ and $\mathbf{n_{+,\times}}$ are column vectors of signal and noise respectively. The noise covariance matrix is:
\begin{equation}
\mathbf{\Sigma}_{n}=\bb \mathbf{\Sigma}_{++} & \mathbf{\Sigma}_{+\times} \\ \mathbf{\Sigma}_{\times+} & \mathbf{\Sigma}_{\times\times} \eb,
\end{equation}
where $\mathbf{\Sigma}_{kk}=\langle\mathbf{{n}}_{k}\mathbf{{n}}_{k}^{T}\rangle$ with the index $k\in \{+, \times\}$ and the brackets $\langle ... \rangle$ denote the ensemble average of the random noise process. (It is understood that $\mathbf{\Sigma}_{++}=\mathbf{\Sigma}_{++}^{T}$, $\mathbf{\Sigma}_{+\times}=\mathbf{\Sigma}_{\times+}^{T}$ and $\mathbf{\Sigma}_{\times\times}=\mathbf{\Sigma}_{\times\times}^{T}$.)

We define the noise-weighted scalar product of two time vectors $\mathbf{x}$ and $\mathbf{y}$ using the noise covariance matrix $\mathbf{\Sigma}_{n}$ as:
\begin{equation}
\left( \mathbf{x}|\mathbf{y} \right)=\mathbf{x}^{T}\mathbf{\Sigma}_{n}^{-1}\mathbf{y}.
\label{scalarP}
\end{equation}
Then the log likelihood ratio $\ln \Lambda$, the logarithmic ratio between the likelihood of the data $\mathbf{d}$ given there is a signal of the form $\mathbf{s}$ present and the likelihood of the data $\mathbf{d}$ given the absence of signals, can be conveniently written as \citep[when assuming Gaussian noise, see e.g.,][for details]{GWDA_LRR12,Ellis2012}:
\begin{equation}
\label{LogLikelihood}
\ln \, \Lambda =\ln \,\frac{\mathcal{L}(\mathbf{s}|\mathbf{d})}{\mathcal{L}(0|\mathbf{d})} =(\mathbf{d}|{\mathbf{s}})-\frac{1}{2}({\mathbf{s}}|{\mathbf{s}}).
\end{equation}
It is worth noting that equation (\ref{LogLikelihood}) is essentially correlating data $\mathbf{d}$ with some signal templates $\mathbf{s}$ and comparing the outputs to a threshold. Regarding $\mathbf{s}$ we rewrite equations (\ref{Ap}-\ref{Ac}) in a more straightforward form:
\begin{equation}
\label{template}
\mathbf{s}_{+,\times}(t)=a_{+,\times}\cos\Phi(t)+b_{+,\times}\sin\Phi(t),
\end{equation}
where the four amplitude parameters are related to physical parameters through:
\begin{equation}
\begin{split}
a_{+}&=\frac{h_0}{2\pi f} [(1+\cos ^{2}\iota) \cos2\psi \sin\Phi_0+2\cos\iota \sin2\psi \cos\Phi_0] \\
b_{+}&=\frac{h_0}{2\pi f} [(1+\cos ^{2}\iota) \cos2\psi \cos\Phi_0-2\cos\iota \sin2\psi \sin\Phi_0] \\
a_{\times}&=\frac{h_0}{2\pi f} [(1+\cos ^{2}\iota) \sin2\psi \sin\Phi_0-2\cos\iota \cos2\psi \cos\Phi_0] \\
b_{\times}&=\frac{h_0}{2\pi f} [(1+\cos ^{2}\iota) \sin2\psi \cos\Phi_0+2\cos\iota \cos2\psi \sin\Phi_0] .
\end{split}
\end{equation}

\begin{figure}
\begin{center}
\includegraphics[width=0.45\textwidth]{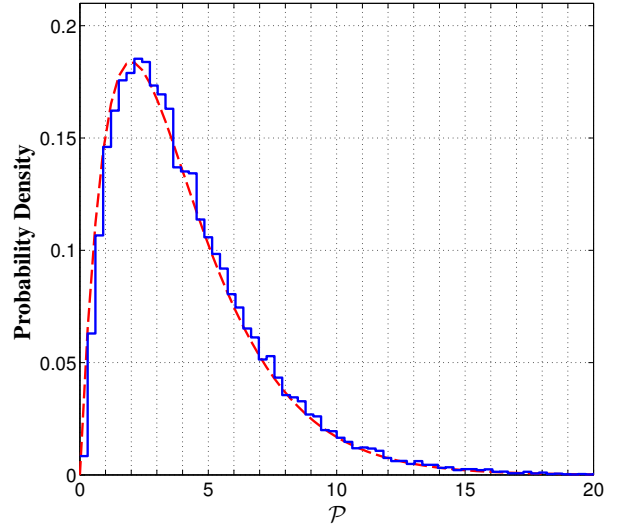}
\end{center}
\caption{Probability distribution of the detection statistics $\mathcal{P}$ in simulated noise-only data (solid blue). We perform 1000 noise realizations using noise models estimated for the PPTA DR1 data and for each realization we search over the same sky direction and 36 independent frequencies. Also shown is the expected $\chi^2$ distribution with four degrees of freedom (red dash).}
\label{fig:PD_DSsim}
\end{figure}

If we write the inverse of the noise covariance matrix in the form of a block matrix:
\begin{equation}
\mathbf{\Sigma}_{n}^{-1}=\bb \mathbf{S}_{11} & \mathbf{S}_{12} \\ \mathbf{S}_{21} & \mathbf{S}_{22} \eb,
\label{Block_Sign}
\end{equation}
and define two column vectors $\mathbf{x}=\cos\Phi(t)$ and $\mathbf{y}=\sin\Phi(t)$, then the maximum likelihood estimators ($\partial \ln \Lambda/\partial a_{+,\times}=0$ and $\partial \ln \Lambda/\partial b_{+,\times}=0$) for the four amplitudes are obtained by solving the following linear equation:
\begin{eqnarray}
 &&\bb \mathbf{d_{+}^{T}}\mathbf{S}_{11}\mathbf{x} + \mathbf{d_{\times}^{T}}\mathbf{S}_{21}\mathbf{x} \\ \mathbf{d_{+}^{T}}\mathbf{S}_{11}\mathbf{y} + \mathbf{d_{\times}^{T}}\mathbf{S}_{21}\mathbf{y} \\ \mathbf{d_{+}^{T}}\mathbf{S}_{12}\mathbf{x} + \mathbf{d_{\times}^{T}}\mathbf{S}_{22}\mathbf{x} \\ \mathbf{d_{+}^{T}}\mathbf{S}_{12}\mathbf{y} + \mathbf{d_{\times}^{T}}\mathbf{S}_{22}\mathbf{y} \eb = \nonumber\\&& \bb \mathbf{x}^{T}\mathbf{S}_{11}\mathbf{x} & \mathbf{x}^{T}\mathbf{S}_{11}\mathbf{y} & \mathbf{x}^{T}\mathbf{S}_{12}\mathbf{x} & \mathbf{x}^{T}\mathbf{S}_{12}\mathbf{y} \\ \mathbf{x}^{T}\mathbf{S}_{11}\mathbf{y} & \mathbf{y}^{T}\mathbf{S}_{11}\mathbf{y} & \mathbf{x}^{T}\mathbf{S}_{21}\mathbf{y} & \mathbf{y}^{T}\mathbf{S}_{21}\mathbf{y} \\ \mathbf{x}^{T}\mathbf{S}_{12}\mathbf{x} & \mathbf{x}^{T}\mathbf{S}_{21}\mathbf{y} & \mathbf{x}^{T}\mathbf{S}_{22}\mathbf{x} & \mathbf{x}^{T}\mathbf{S}_{22}\mathbf{y} \\ \mathbf{x}^{T}\mathbf{S}_{12}\mathbf{y} & \mathbf{y}^{T}\mathbf{S}_{21}\mathbf{y} & \mathbf{x}^{T}\mathbf{S}_{22}\mathbf{y} & \mathbf{y}^{T}\mathbf{S}_{22}\mathbf{y} \eb \bb \hat{a}_{+} \\ \hat{b}_{+} \\ \hat{a}_{\times} \\ \hat{b}_{\times} \eb,
 \label{MLE4Amp}
\end{eqnarray}
where we have applied the properties of $\mathbf{S}_{11}=\mathbf{S}_{11}^{T}$, $\mathbf{S}_{21}=\mathbf{S}_{12}^{T}$ and $\mathbf{S}_{22}=\mathbf{S}_{22}^{T}$. Estimates of $a_{+,\times}$ and $b_{+,\times}$ obtained by solving equation (\ref{MLE4Amp}) are used to calculate the log likelihood ratio with equations (\ref{LogLikelihood}-\ref{template}). Defining a column vector $\mathbf{\lambda}$ that contains maximum likelihood estimates of the four amplitudes $[\hat{a}_{+};\hat{b}_{+};\hat{a}_{\times};\hat{b}_{\times}]$ and multiplying $\mathbf{\lambda}^{T}$ to both sides of equation (\ref{MLE4Amp}), one finds that $(\mathbf{d}|{\mathbf{s}})=({\mathbf{s}}|{\mathbf{s}})$. Our detection statistic is taken as $\mathcal{P}=2\ln \Lambda$ and thus the signal-to-noise ratio as defined by $\rho = \sqrt{({\mathbf{s}}|{\mathbf{s}})}$ is related to $\mathcal{P}$ through $\rho=\sqrt{\mathcal{P}}$. Note that $\mathcal{P}$ in the case of Gaussian noise-only data follows a $\chi^2$ distribution with four degrees of freedom. Fig. \ref{fig:PD_DSsim} shows the probability distribution of $\mathcal{P}$ in simulated noise-only data as compared against the expected distribution.

It is worth mentioning that the derivation of $\mathcal{P}$ follows that of the $\mathcal{F}$-statistic in the context of CW search using ground-based interferometers \citep{Fsts98}. The $\mathcal{F}$-statistic was also adapted in \citet{Ellis2012} for PTA data to derive the coherent $\mathcal{F}_{e}$-statistic and incoherent $\mathcal{F}_{p}$-statistic (equivalent to a power spectral summing technique). Our statistic is similar to the $\mathcal{F}_{e}$-statistic in the following ways: 1) it is implemented fully in the time domain; 2) it targets at the coherent Earth-term signals; 3) it has been maximized over extrinsic amplitude parameters. The main difference is that the $\mathcal{F}_{e}$-statistic applies directly on timing residual data whereas $\mathcal{P}$ works with the reduced $A_{+,\times}(t)$ data.

\subsubsection{Maximum likelihood problem in degenerate multivariate Gaussian noise}
\label{ML-MVN}
The constraints set on $A_{+,\times}(t)$ result in some degeneracy in the data, and therefore their covariance matrix $\mathbf{\Sigma}_{n}$ is no longer a full-rank matrix. Specifically the rank $r$ of $\mathbf{\Sigma}_{n}$ is given by $n-m$ with $n$ the total number of data points in the stacked $[\mathbf{A}_{+};\mathbf{A}_{\times}]$ data and $m$ the number of constraints that have been applied.

The above problem corresponds to finding the maximum likelihood solution in the case of degenerate multivariate Gaussian noise. One needs to replace $\mathbf{\Sigma}_{n}^{-1}$ in equations (\ref{scalarP}) and (\ref{Block_Sign}) with the generalized inverse which can be obtained via eigen-decomposition ($\mathbf{\Sigma}_{n}= \mathbf{E} \mathbf{D} \mathbf{E}^{T}$):
\begin{equation}
\mathbf{\Sigma}_{n}^{-}= \mathbf{E} \mathbf{D}^{-1} \mathbf{E}^{T} \,\ ,
\label{InvSign}
\end{equation}
where $\mathbf{E}$ is $n \times r$ of full rank $r$, and has columns that are the eigenvectors corresponding to the positive eigenvalues of $\mathbf{\Sigma}_{n}$ as given in the diagonal elements of the $r \times r$ diagonal matrix $\mathbf{D}$. It is straightforward to show that $\mathbf{\Sigma}_{n}= \mathbf{U} \mathbf{U}^{T}$ where $\mathbf{U}=\mathbf{E}\sqrt{\mathbf{D}}$ and we can use $\mathbf{U}_{\rm{left}}^{-1}$ (defined as $\mathbf{U}_{\rm{left}}^{-1}\mathbf{U}=\mathbf{I}_{r}$ with $\mathbf{I}_{r}$ being the $r \times r$ identity matrix) to whiten the $A_{+,\times}(t)$ time series. If we write $\mathbf{A}=[\mathbf{A}_{+};\mathbf{A}_{\times}]$, then elements in $\mathbf{A}_{w}=\mathbf{U}_{\rm{left}}^{-1}\mathbf{A}$ independently follow the standard Gaussian distribution.

\subsection{Detection and bounding techniques}
\label{DetPipe}
In a detection problem, we are concerned with the false alarm probability (FAP) of a measured detection statistic $\mathcal{P}_{\rm{obs}}$, which is the probability that $\mathcal{P}$ exceeds $\mathcal{P}_{\rm{obs}}$ for noise-only data. So the single-trial FAP is given by $1-\mathrm{CDF}(\mathcal{P}_{\rm{obs}};\chi_{4}^{2})$ where $\mathrm{CDF}(\, ;\chi_{4}^{2})$ denotes the cumulative distribution function (CDF) for a $\chi^2$ distribution with four degrees of freedom. Without prior knowledge of GW frequency and source sky location, a search should be performed in the 3-dimensional parameter space ($\delta$, $\alpha$, $f$). This introduces a trials factor $N_{\rm{trial}}$ defined as the number of independent cells in the searched parameter space. We are interested in the total FAP as given by $1-[\mathrm{CDF}(\mathcal{P}_{\rm{max}};\chi_{4}^{2})]^{N_{\rm{trial}}}$ for the maximum detection statistic $\mathcal{P}_{\rm{max}}$ found in the search.

In a standard \textsc{TEMPO2} generalized least-squares fit, in addition to a full timing model for each pulsar we globally fit for $A_{+,\times}(t)$ for evenly spaced times between MJD 53430 and 55620 with a sampling interval of 30 days for a set of sky positions (with the number determined below). The sampling we choose for the $A_{+,\times}(t)$ fit is slightly lower than our observing cadence to ensure that there are $\gtrsim 20$ observations for each sampled epoch. It is well understood that for evenly sampled data independent frequencies are defined as a set of harmonics of $\Delta f=1/T_{\rm{span}}$ (with $T_{\rm{span}}$ being the data span) up to the Nyquist frequency. Determined by the sampling for the $A_{+,\times}$ fit, the frequency range of our search $5\times 10^{-9}$--$2\times 10^{-7}$ Hz consists of 36 independent frequency channels (denoted as $N_{f}=36$). In this work we choose to analyze 141 GW frequencies from $0.5\Delta f$ to $36\Delta f$ with an interval of $0.25\Delta f$ (because of the constraints set on $A_{+,\times}(t)$, two frequencies close to 1 $\rm{yr}^{-1}$ are excluded in the analysis).

Regarding the number of sky directions that need to be searched, one should consider that:\newline
1) it must not be too small otherwise one could miss a potential signal due to a mismatch in ($\delta$, $\alpha$);\newline
2) there is an upper limit for the number of independent sky positions (denoted as $N_{\rm{sky}}$) -- the PPTA DR1 data set contains about 4000 ToAs (i.e., $N_{\rm{trial}}<4000$), which implies $N_{\rm{sky}}\lesssim 100$ if we assume $N_{\rm{trial}}=N_{f} N_{\rm{sky}}$ by neglecting the correlation between GW frequency and source location\footnote{A strong correlation exists between GW source location and frequency at low frequencies where sources are essentially non-evolving. However, as the search targets at only the Earth terms and we are in a weak-signal limit the correlation is not significant.}. Note that the purpose of this assumption is only to have a rough estimate of $N_{\rm{sky}}$ which is used in determining the detection threshold for all-sky sensitivities.\newline
For the present search we choose to use an uniform sky grid consisting of 1000 points. We find that using a finer sky grid of 4000 points results in $\ll 1\%$ increase in the maximum detection statistic for the PPTA DR1 data set. (However, for better visual quality of the figures sky maps shown in the paper consist of 4000 pixels.)

We set a total FAP of $1\%$ as the detection threshold for the all-sky blind search. Because $N_{\rm{trial}}$ is not precisely known, we choose to estimate the FAP for the most significant ``event" found in the search by simulations as later described in section \ref{SearchResults}.

Our steps towards detection are as follows:
\begin{enumerate}
\item For each grid point on the sky, we use the \textsc{TEMPO2} software package to simultaneously determine timing parameters for a full timing model of 20 PPTA pulsars and to form estimates of $A_{+,\times}(t)$ and their covariance matrix $\mathbf{\Sigma}_{n}$ through a global generalized least-squares fit.

\item For each frequency $f_{j}$, we calculate and record values of ``observed" detection statistic $\mathcal{P}_{\rm{obs}}(f_{j})$. The first two steps can be accomplished using the \textsc{TEMPO2} \textit{findCWs} plugin.

\item Repeat steps 1-2 for other sky directions. If anywhere we find a $\mathcal{P}_{\rm{obs}}$ that corresponds to a FAP of less than $1\%$, we claim a detection.
\end{enumerate}

In the absence of a detection, our steps towards setting upper limits on $h_{0}$ as a function of GW frequencies $f$ are as follows:
\begin{enumerate}
\item At each frequency $f_{j}$, we generate SMBBH signals for a given value of $h_0$ and other parameters ($\cos\delta$, $\alpha$, $\cos\iota$, $\psi$ and $\Phi_{0}$) randomly chosen from uniform distributions over their applicable ranges and add such signals to simulated noise. For completeness we also consider the case that CW signals are injected to real data.

\item The detection statistic is evaluated at the injected source sky location\footnote{It could have been more appropriate to search over a sky patch centred around the randomly chosen sky location. We choose to search only at the injected source position because of limited computing resources.} and frequency, and recorded as the ``simulated" detection statistic $\mathcal{P}_{\rm{sim}}(f_{j})$, which is then compared against $\mathcal{P}_{\rm{obs}}(f_{j})$ found in the same frequency and sky position.

\item Perform 1000 simulations with each having source parameters randomized and independent noise realizations, and adjust $h_0$ until $95\%$ of the injections lead to a value of $\mathcal{P}_{\rm{sim}}(f_{j})> \mathcal{P}_{\rm{obs}}(f_{j})$. Record the value of $h_0(f_{j})$ as the frequentist $95\%$ confidence upper limit. Since we use 1000 simulations, the 1--$\sigma$ uncertainty of the confidence is only $0.1\%$ given a binomial statistic.
\end{enumerate}

While upper limits tell us the maximum amplitude of a GW at a particular frequency that is consistent with our observations, we are also interested in the sensitivities achieved by the PPTA DR1 data set, i.e., the minimum values of $h_0$ that would produce a detectable signal in our data set with $95\%$ probability. The procedure of producing an all-sky sensitivity curve is the same as that for bounding as outlined above except that $\mathcal{P}_{\rm{sim}}$ should now be compared against a fixed detection threshold. For all-sky sensitivities, the detection threshold is chosen as $\mathcal{P}_{\rm{th,all-sky}}=20.9$ using $N_{\rm{sky}}=30$ -- the reduced trials factor as we have specified GW frequencies to obtain sensitivities. This rough estimate of $N_{\rm{sky}}$ is simply taken as $N_{\rm{trial}}/36$ and we consider $N_{\rm{trial}}=1000$ for the PPTA DR1 data set as estimated through simulations in section \ref{SearchResults}.

As it is expected that the sensitivity varies significantly across the sky, we additionally present in this paper directional sensitivity curves for the most, median and least sensitive sky direction given the PPTA DR1 data set. In this case the detection threshold is $\mathcal{P}_{\rm{th,direc}}=23.5$ corresponding to a single-trial FAP of $10^{-4}$. This is appropriate for the case that in a targeted search, the source sky location and frequency are known, e.g., for SMBBH candidates from electromagnetic observations \citep[see, e.g.,][and references therein]{SarahCQG13}. Usually in a directional search the source orbital frequency is unknown, e.g., for the so-called GW hotspots \citep{Rosado14,SimonGWhot14}. In this case the threshold should be increased to $31.2$ to account for a trials factor of $N_{f}=36$ and the corresponding sensitivities would be decreased by $15\%$. Because of computational limitations the process for obtaining directional sensitivities is simplified as follows:\newline
1) At each frequency and each one of the 1000 sky grid points we inject to the PPTA DR1 data CW signals with a fixed $h_0$, $\cos \iota=1$, and random values for $\psi$ and $\Phi_{0}$ over $[0,2\pi)$. Then we calculate the signal-to-noise ratio $\rho$;\newline
2) Because for a given noise realization (in our case the PPTA DR1 data), $\rho$ scales linearly as $h_0$, values of $\rho$ obtained in the previous step can be scaled to the given detection threshold to obtain detection sensitivities on $h_0$. Here a multiplying factor of $\sqrt{5/2}$ is included to account for average binary orientations. This factor corresponds to the difference in the square root between the maximum value (two) for $[(1+\cos ^{2}\iota)/2]^{2}+\cos ^{2}\iota$ and its average value ($4/5$) for an uniform distribution of $\cos \iota$ in $(-1,1]$. \newline
Using the above method we can also generate a sky map of sensitivities at a given frequency. Note that the value of injected $h_0$ does not matter as long as it is large enough so that the injection at the least sensitive sky direction results in a detection. The purpose of these directional sensitivity plots is just to show what sensitivity is available with the current data set for directional/targeted searches.

\section{RESULTS AND DISCUSSION}
\label{sec:results}
In this section, we present verification of the data analysis pipeline, followed by search results, upper limits and sensitivities based on the PPTA DR1 data set.
\subsection{Verification of the pipeline}
\begin{figure*}
\centerline{\includegraphics[width=14cm]{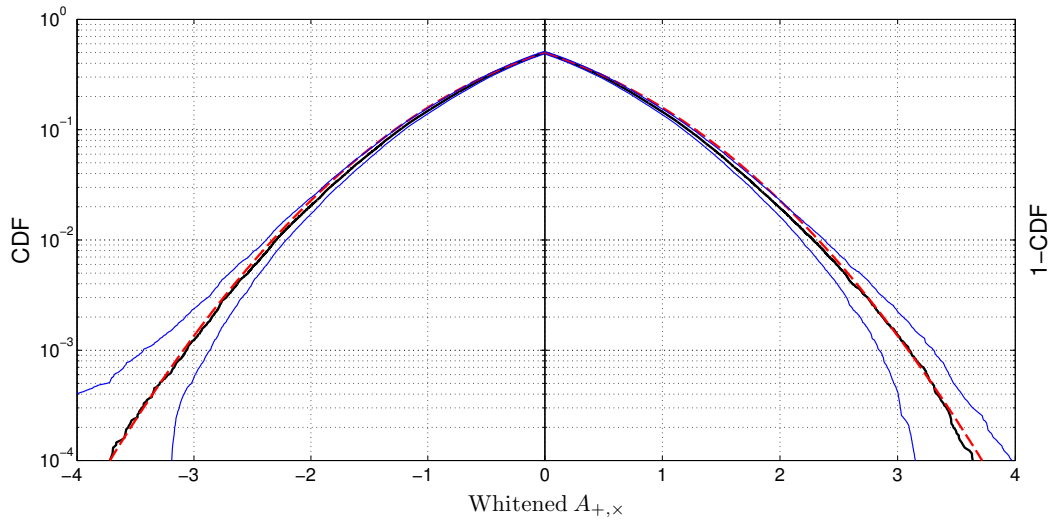}}
\caption{
\label{fig:ApcWhite}
Empirical CDF (thick solid black) and its 2--$\sigma$ confidence region (thin solid blue) for the whitened $A_{+,\times}(t)$ data obtained from the PPTA DR1 data set, compared against the standard Gaussian distribution (red dash).}
\end{figure*}

The analysis presented here has undergone extensive checking to make sure the pipeline works as expected. The first test is to confirm that the whitened $A_{+,\times}(t)$ data follow the standard Gaussian distribution. For each searched sky position, we whiten the derived $A_{+,\times}(t)$ time series using the noise covariance matrix as described in section \ref{ML-MVN}. Fig. \ref{fig:ApcWhite} shows the empirical CDF and its 2--$\sigma$ confidence region for the whitened $A_{+,\times}(t)$ data, which agrees well with the standard Gaussian distribution.

\begin{figure}
\includegraphics[width=0.48\textwidth]{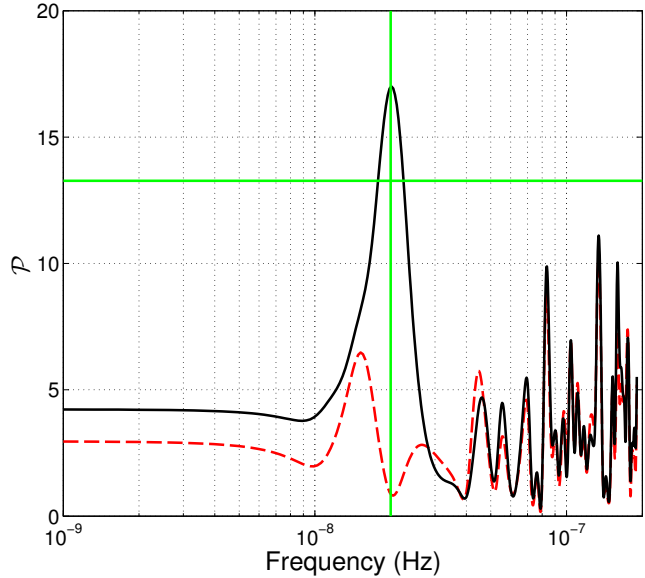}
\caption{Detection statistics ($\mathcal{P}$) as a function of frequencies for the PPTA DR1 data set (red dash) and for the same data set but has included a signal injection (solid black). Both were evaluated at the sky location of the Virgo cluster where the signal was injected. The vertical line marks the injected frequency (20 nHz), while the horizontal line corresponds to a single-trial FAP of $1\%$.}
\label{fig:DStestVirgo}
\end{figure}

The second verification process involves the correct reconstruction of signal injections. In Fig. \ref{fig:DStestVirgo}, we show the detection statistics as a function of frequencies in the case of a CW signal specified with the following parameters: $M_{c}=7\times 10^{8} M_{\odot}$, $d_{L}=16.5$ Mpc, $f=20$ nHz, $\cos\iota=0.5$, $\psi=\Phi_{0}=0$, $\alpha=3.2594$ and $\delta=0.2219$ (the sky location of the Virgo cluster) was injected to the PPTA DR1 data set. For this case the search was performed at the injected sky location and we use a frequency interval of $\Delta f/16$ to increase frequency resolution. The maximum detection statistic was found at a frequency of $20.15$ nHz. As a comparison, we also show the detection statistics measured for the real data set.

\begin{figure*}
\begin{center}
        \subfigure[]{%
            \label{fig:sim1}
            \includegraphics[width=0.58\textwidth]{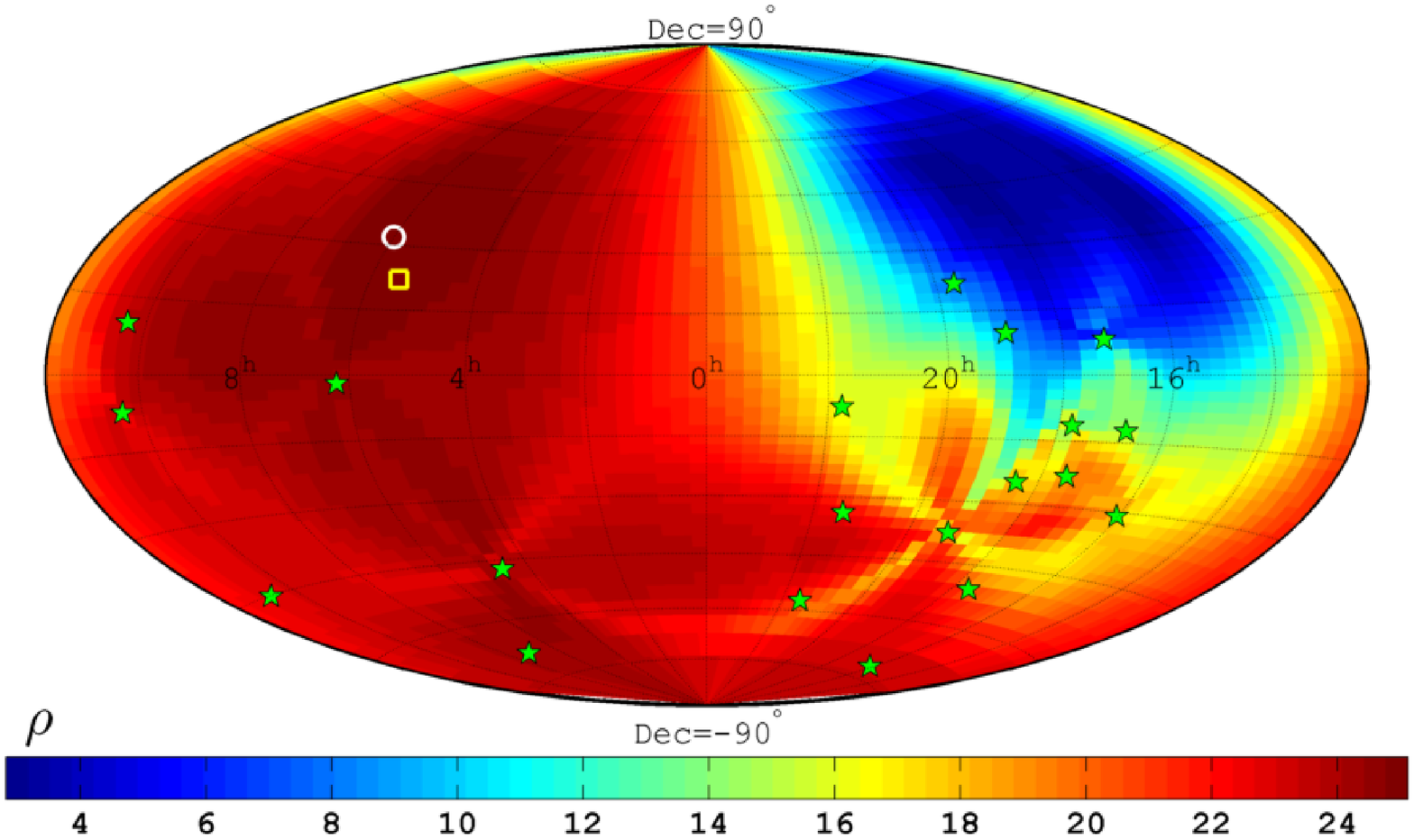}
        }\\
        \vspace{-3mm}
        \subfigure[]{%
            \label{fig:sim2}
            \includegraphics[width=0.58\textwidth]{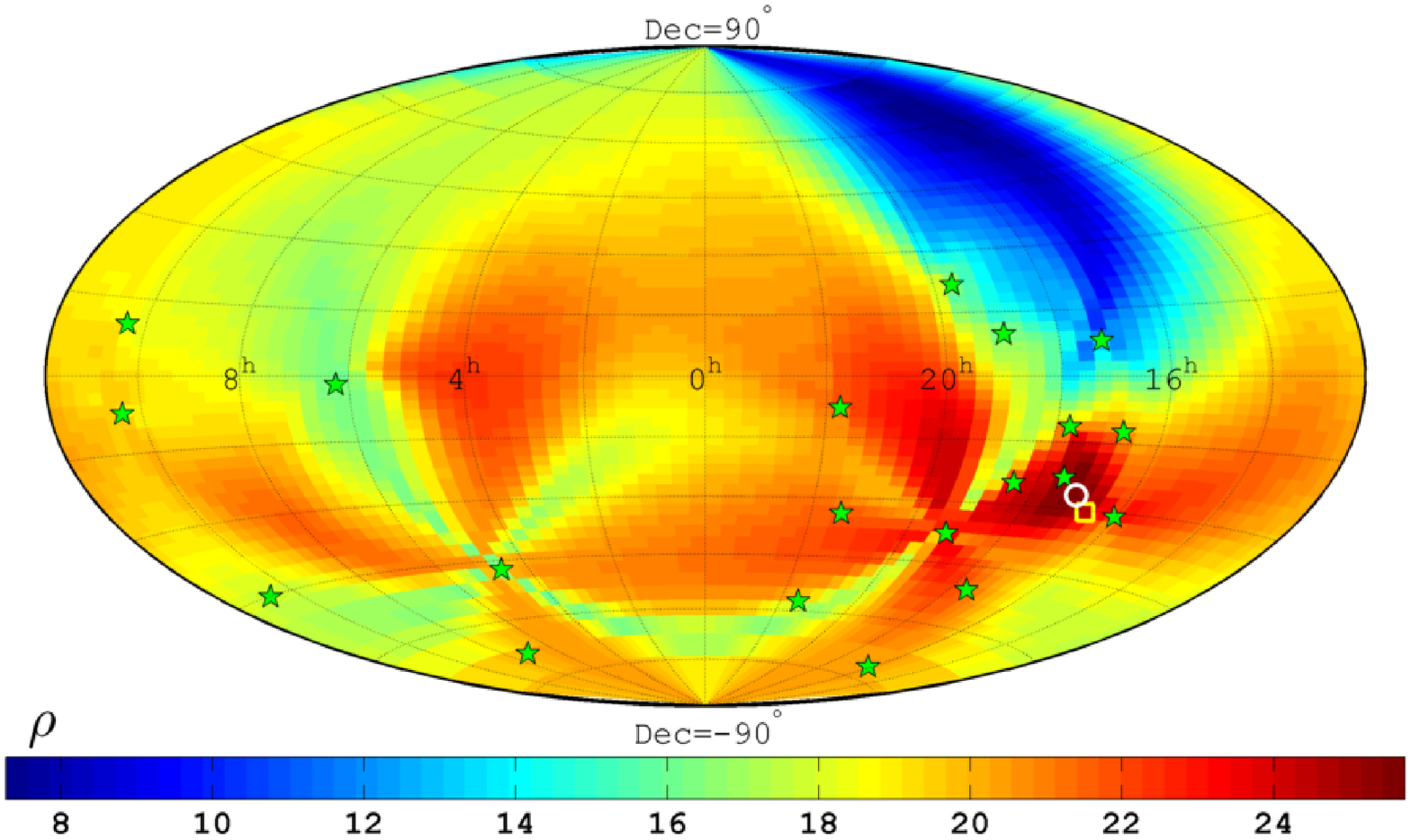}
        }
\end{center}
\caption{Sky map of signal-to-noise ratios ($\rho$) for simulated data set that includes a strong signal injection made in the least (a) or most (b) sensitive sky region. The signal is injected at the location indicated by a ``$\square$" and the maximum $\rho$ is found at ``$\circ$". Sky locations of the 20 PPTA pulsars are marked with ``$\star$".}
\label{fig:sim}
\end{figure*}

Fig. \ref{fig:sim} shows sky maps of signal-to-noise ratios ($\rho$) for two strong signal injections made to simulated noise. The source sky locations were chosen in the least and most sensitive sky region, as will be illustrated later in Fig. \ref{fig:SenSkyF2}. In both cases the maximum $\rho$ is found at a grid point near the injected sky location, but the error box varies dramatically (a few tens square degrees in comparison to thousands of square degrees).

\begin{figure}
\includegraphics[width=0.48\textwidth]{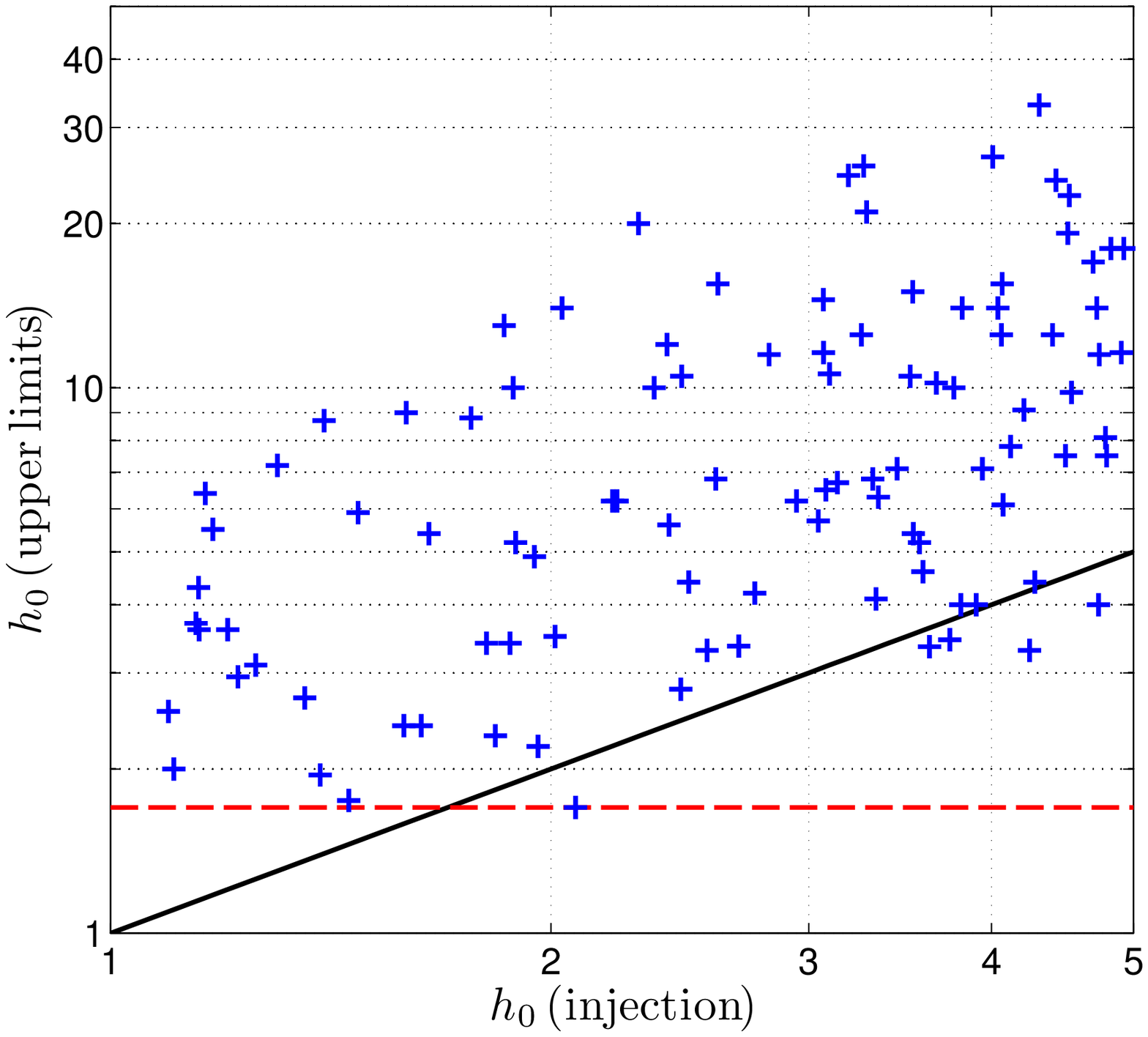}
\caption{Upper limits on $h_0$ established for real data with signal injections, are compared against injected values (both in $10^{-14}$). Each point represents a separate injection. The horizontal dash line marks the PPTA DR1 upper limit ($1.7\times 10^{-14}$) at $10^{-8}$ Hz where all injections are made. The solid black line is for upper limits equal the injected values.}
\label{fig:h0LimInj}
\end{figure}

The final test is that for a data set that contains an injected CW signal, the established upper limit on $h_0$ should be above the injected value. Fig. \ref{fig:h0LimInj} shows results of such a test for 100 signal injections made to real data. All signals are simulated at $10^{-8}$ Hz with $h_0$ uniformly distributed in 1--$5 \times 10^{-14}$ and other parameters randomly chosen from uniform distributions over their applicable ranges. The PPTA DR1 upper limit at this frequency is $1.7 \times 10^{-14}$. Note that 5 out of 100 injections have failed this test, which is expected as the upper limits are at a $95\%$ confidence level.
\begin{figure*}
\centerline{\includegraphics[width=12cm]{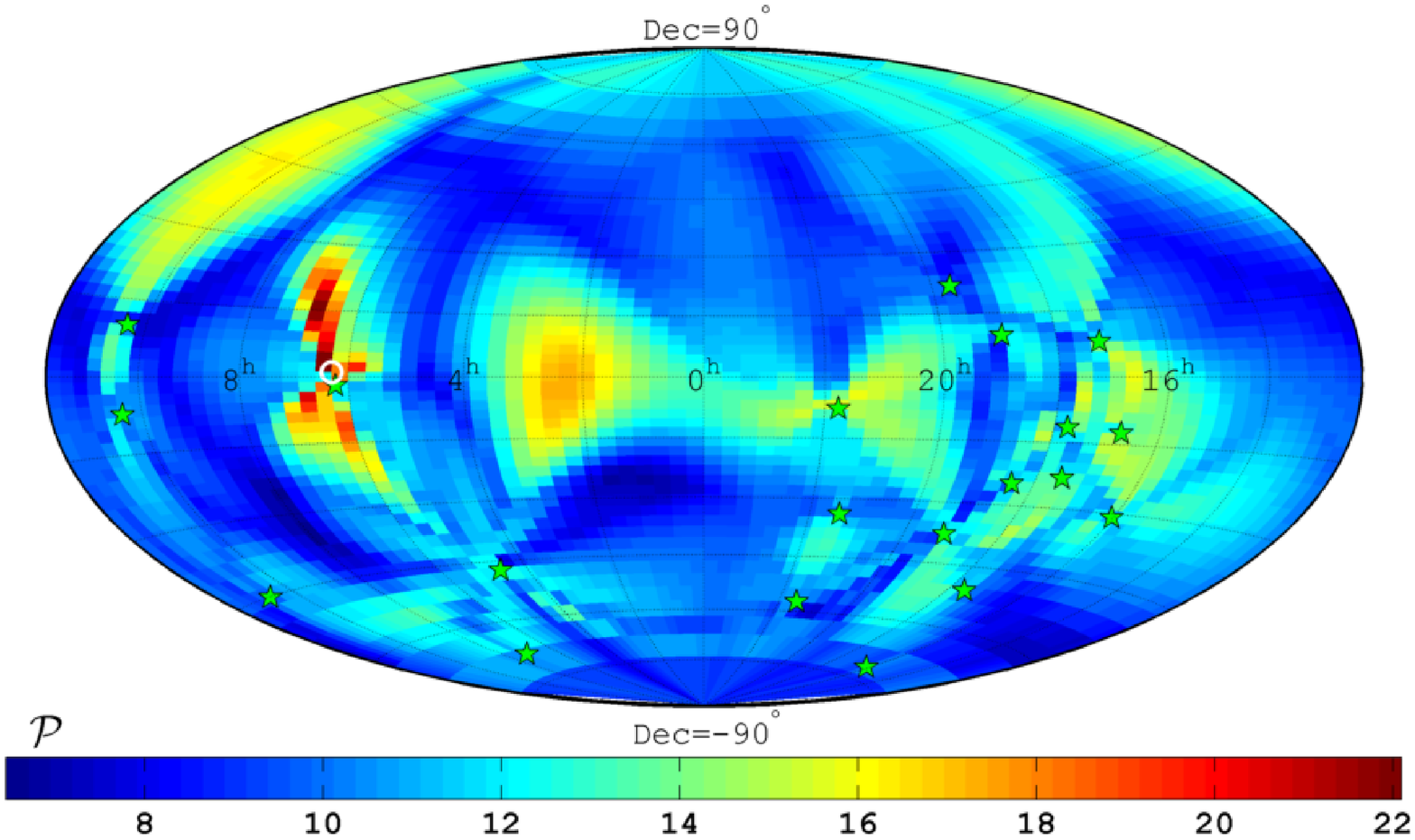}}
\caption{Sky map of the detection statistics ($\mathcal{P}$) measured for the PPTA DR1 data set. The most significant statistic is found at a direction indicated by ``$\circ$". Sky locations of the 20 PPTA pulsars are labeled with ``$\star$".}
\label{fig:det}
\end{figure*}

\begin{figure}
\includegraphics[width=0.48\textwidth]{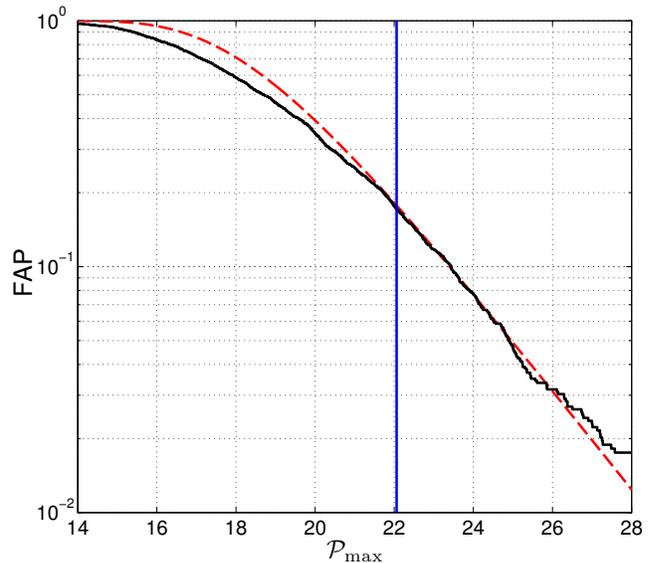}
\caption{FAP as a function of the maximum detection statistic ($\mathcal{P}_{\rm{max}}$) as determined by 1000 simulations (solid black). The red dash line is for a $\chi^2$ distribution with four degrees of freedom assuming a trials factor $N_{\rm{trial}}=1000$. The vertical line marks $\mathcal{P}_{\rm{max}}=22.06$ measured for the PPTA DR1 data set.}
\label{fig:FAP}
\end{figure}

\subsection{Search results}
\label{SearchResults}
We show in Fig. \ref{fig:det} a sky map of the detection statistic maximized over 36 frequency channels for each sky direction. The most significant value $\mathcal{P}_{\rm{max}}=22.06$ across the sky is found at $85$ nHz. To estimate the FAP of this ``event", we produce simulated data sets and treat them exactly the same as real data set, i.e., going through the same fitting process and searching over exactly the same grid points in the parameter space. For each noise realization we record the maximum value of the detection statistic. With 1000 simulations we show in Fig. \ref{fig:FAP} FAP estimates based on the empirical distribution of $\mathcal{P}_{\rm{max}}$. The FAP for $\mathcal{P}_{\rm{max}}=22.06$ found in the PPTA DR1 data set is estimated to be $17\%$, implying the search result is consistent with a null detection. From the empirical distribution of $\mathcal{P}_{\rm{max}}$, we also obtain an estimate of $N_{\rm{trial}}=1000$ for the PPTA DR1 data set.

\subsection{Upper limits and sensitivities}
\label{LimSen}
\begin{figure}
\begin{center}
\includegraphics[width=0.46\textwidth]{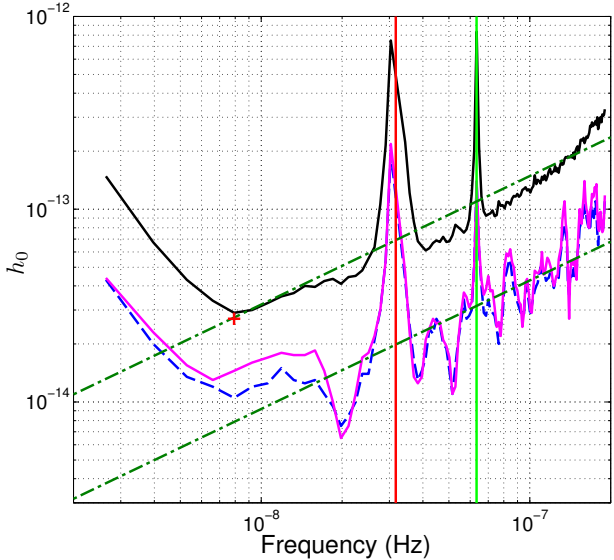}
\end{center}
\caption{All-sky upper limits on $h_0$ as a function of GW frequencies for two cases: signals injected to real data (dash blue) or simulated noise (solid pink). The all-sky sensitivity curve (solid black) is obtained for simulated noise. Two vertical lines correspond to frequencies of 1 and 2 $\rm{yr}^{-1}$. The dash-dotted straight lines are strain amplitudes expected from SMBBHs with $M_{c}=10^{10} M_{\odot}$ and $d_{L}=400$ Mpc (upper), or $M_{c}=10^{9} M_{\odot}$ and $d_{L}=30$ Mpc (lower). The point marked by a plus sign is the sensitivity calculated for evolving sources assuming $M_{c}=10^{10} M_{\odot}$.}
\label{fig:Limits}
\end{figure}

Fig. \ref{fig:Limits} shows all-sky upper limits on $h_0$ for two cases: CW signals are injected to a) real data or b) simulated noise. One can see that case b) gives slightly worse upper limits across the whole frequency band (most notably between 5 and 15 nHz) and the noisy trend in frequency for both set of limits is almost identical. Also shown in Fig. \ref{fig:Limits} is an all-sky sensitivity curve for case b). This sensitivity curve is roughly a factor of two above the upper limit curve. This is expected because in the process of setting an upper limit $\mathcal{P}_{\rm{sim}}$ is compared against $\mathcal{P}_{\rm{obs}}$ which has an average of four (the distribution of $\mathcal{P}_{\rm{obs}}$ varies significantly over frequency, resulting in the noisiness in the upper limit curves), and the threshold $\mathcal{P}_{\rm{th,all-sky}}=20.877$ is a factor of five higher than $\mathcal{P}_{\rm{obs}}$ on average (taking $\sqrt{5}$ for scaling in $h_0$). In order to test the effect of our assumption that pulsar terms are in the same frequency bin as Earth terms, we calculate the sensitivity at the 5th bin (the most sensitive bin, with a centre frequency 8 nHz) for evolving sources with $M_{c}=10^{10} M_{\odot}$. As indicated by the ``plus" sign in Fig. \ref{fig:Limits}, the sensitivity at this frequency bin is only increased by $7\%$.

\begin{figure}
\begin{center}
\includegraphics[width=0.46\textwidth]{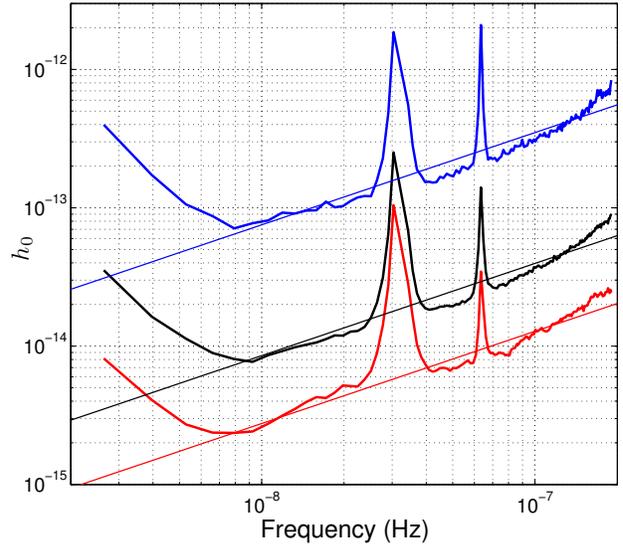}
\end{center}
\caption{Sensitivities as a function of GW frequencies for the most (lower), median (middle) and least (upper) sensitive sky direction given the PPTA DR1 data. Straight lines are for SMBBHs with $M_{c}=10^{9} M_{\odot}$ and $d_{L}=100$ Mpc (lower), $M_{c}=10^{10} M_{\odot}$ and $d_{L}=1.5$ Gpc (middle), $M_{c}=10^{10} M_{\odot}$ and $d_{L}=170$ Mpc (upper), that could produce CW signals at the level of the three sensitivity curves between $10^{-8}$ and $10^{-7}$ Hz (except two narrow bands centred around 1 and 2 $\rm{yr}^{-1}$).}
\label{fig:SenSky}
\end{figure}

Our upper limits are about a factor of four better than the previously published limits in PPTA10. This improvement is mainly because the new data set has significantly improved timing precision and cadence over the earlier data set. Our limits also improve by a factor of two on those reported in the recent paper by NANOGrav \citep{NANOGravCW14}, comparing to their results based on ``fixed-noise" approaches. This improvement is mostly caused by the higher observing cadence, the slightly longer data span and the much larger number of independent observing sessions in the PPTA data set. It should be noted that: 1) there is a factor of $\sqrt{8/5}$ difference in the definition of the GW strain amplitude being constrained in PPTA10. Their upper limits were set on the inclination-averaged mean-square amplitude that is given by $h_{0}\times \sqrt{8/5}$ \citep[see, e.g., equation (24) in][]{Jaffe_Backer03}; 2) As CW signals were under-represented in PPTA10 by a factor of $\sqrt{2}$, corresponding to the difference between the maximum amplitude of a sinusoid and its rms amplitude, we have divided upper limits presented in that paper by the same factor when making comparisons.

\begin{figure*}
\centerline{\includegraphics[width=12cm]{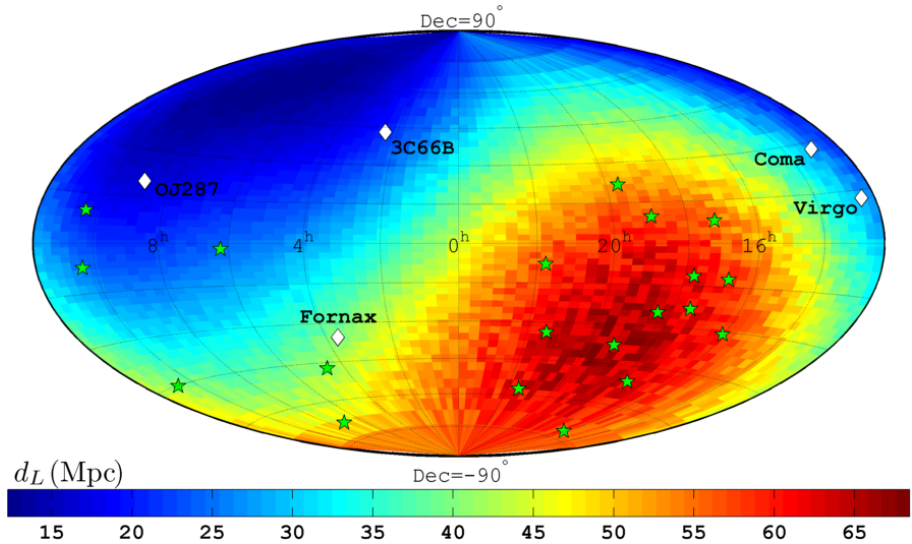}}
\caption{Sky map of luminosity distance ($d_L$) out to which the PPTA DR1 data set is sensitive at $10^{-8}$ Hz to CW signals from circular SMBBHs of chirp masses $10^{9} M_{\odot}$. Sky locations of the 20 PPTA pulsars are labeled with ``$\star$". White diamonds mark the location of possible SMBBH candidates or nearby clusters (luminosity distances are 92 Mpc, 102 Mpc, 19 Mpc, 16.5 Mpc and 1.07 Gpc for 3C66B, Coma, Fornax, Virgo and OJ287 respectively). Note that the sensitivity in $d_L$ scales as $M_{c}^{5/3}$ and is roughly the same for the frequency band of $10^{-8}$--$10^{-7}$ Hz except two narrow bands centred around 1 and 2 $\rm{yr}^{-1}$ (see Fig. \ref{fig:SenSky}).}
\label{fig:SenSkyF2}
\end{figure*}

In Fig. \ref{fig:SenSky} we show sensitivity curves for the most, median and least sensitive sky direction given the PPTA DR1 data. It should be noted that such sensitivities are to be used as a guide for targeted searches, i.e., for known source sky locations and frequencies. For the most sensitive sky direction, the current data set is sensitive to average-oriented SMBBHs of chirp masses $10^{9} M_{\odot}$ up to about 100 Mpc in the frequency band of $10^{-8}$--$10^{-7}$ Hz (except two narrow bands centred around 1 and 2 $\rm{yr}^{-1}$). For directional searches with unknown orbital frequencies sensitivities would be decreased by $15\%$ to account for the trials factor involved in a search over frequency as discussed in section \ref{DetPipe}. The median sensitivity curve is a factor of four below the all-sky sensitivity curve because for $95\%$ of the whole sky sources with $h_0$ above the latter can be detected while the former only applies to $\sim 50\%$ of the sky. For both Figs \ref{fig:Limits} and \ref{fig:SenSky}, the huge sensitivity loss at around 1 and 2 $\rm{yr}^{-1}$ is due to the constraints set on $A_{+,\times}(t)$ which aim to avoid the covariance between a global fit for $A_{+,\times}(t)$ and the fit for positions, proper motions and parallax of individual pulsars.

Fig. \ref{fig:SenSkyF2} shows the distance at which a circular SMBBH of a certain chirp mass would produce a detectable signal at $10^{-8}$ Hz in our data set. In this plot the signal injections only include Earth terms because the inclusion of pulsar terms bias the sky localization and thus make the sky map very noisy. However, we emphasize that similar results should be obtainable if we search over the sky (rather than only at the injected location) for each signal injection when pulsar terms are included. The purpose of Fig. \ref{fig:SenSkyF2} is just to illustrate the PPTA CW sensitivity map and to gain some insights on how the addition of new pulsars to the timing array helps. As expected, we are most sensitive in the sky region where the best-timed PPTA pulsars are located and least sensitive in the opposite direction. Main findings from Fig. \ref{fig:SenSkyF2} include:\newline
1) The PPTA DR1 data set is sensitive to potential SMBBHs with chirp masses of $\gtrsim 2.3\times 10^{9} M_{\odot}$ in the Coma Cluster, and of $\gtrsim 7\times 10^{8} M_{\odot}$ for both the Fornax Cluster and Virgo Cluster;\newline
2) With the current analysis we are unable to place meaningful constraints on 3C66B as it was proposed to have an orbital period of 1.05 yr \citep{3C66B10ApJ} where our sensitivity is very low because of the biannual sinusoidal constraint set on $A_{+,\times}(t)$ (see Figs \ref{fig:Limits} and \ref{fig:SenSky});\newline
3) For another SMBBH candidate OJ287, modelled with an orbital period of 12 yrs \citep{OJ287ApJ97}, a possible constraint on the chirp mass with the current data set would be $\sim 10^{10} M_{\odot}$ which is about an order of magnitude higher than the current mass estimate \citep{OJ287mass};\newline
4) The possible SMBBH candidates and nearby clusters are all located in the insensitive sky region. This shows the benefits of adding new good pulsars to increase the PPTA's astrophysical reach.
\subsubsection{Upper limits on the SMBBH coalescence rate}
\label{LimRate}
Given the absence of CW signals in the PPTA DR1 data set, upper limits on the coalescence rate of SMBBHs can be computed in a straightforward way. Following \citet{ZLWen11LimRate}, but rather than constraining the differential coalescence rate (with respect to chirp mass and redshift), we wish to set limits on the local coalescence rate density. The expected number of events can be written as $\mu =R \sum_{i}\epsilon V_{i}\Delta T(f_{i})$ where $R$ is the coalescence rate per unit volume, $\epsilon$ is the detection efficiency (which is $95\%$ for all points on the all-sky sensitivity curve shown in Fig. \ref{fig:Limits}), $V_{i}$ is the sensitive volume at frequency $f_{i}$ (simply taken as $4\pi d_{L,i}^{3}/3$ with $d_{L,i}$ being the luminosity distance out to which a SMBBH would produce a detectable CW signal at $f_{i}$ with $95\%$ probability) and $\Delta T(f_{i})$ is the time duration that a binary stays in the $i$-th frequency bin. Assuming Poisson-distributed events, the probability of no events being detected is $e^{-\mu}$. Therefore, the frequentist $95\%$ confidence upper limit is $R_{95\%}=-{\rm{ln}}(1-0.95)/\sum_{i}\epsilon V_{i}\Delta T(f_{i})$.

Making use of the all-sky sensitivities for 141 frequencies shown in Fig. \ref{fig:Limits} together with the assumption that the sensitivity is a constant for each frequency bin with width 1.32 nHz, we find that $R_{95\%} = 4\times 10^{-3} (10^{10} M_{\odot}/M_{c})^{10/3} {\rm{Mpc}}^{-3} {\rm{Gyr}}^{-1}$ for nearby ($z\lesssim 0.1$) SMBBHs (with the maximum applicable redshift corresponding to the largest $d_{L,i}$ for $M_{c}=10^{10} M_{\odot}$). Note that our limit is about two orders of magnitudes above the current estimates of galaxy merger rate density in the local Universe \citep[see, e.g., fig. 13 in][]{ARAA14}.
\section{CONCLUSIONS}
\label{sec:conclu}
Over the past few years, PTAs have been collecting pulsar timing data with steadily improving precision and are starting to set astrophysically interesting upper limits. However, most of the analyses of PTA data have focused on a stochastic background that could be produced by the combined emission by a large number of individual SMBBHs. In this paper we have developed a new coherent method for detection of individual CW sources and have tested it extensively on both simulated and real pulsar timing data. The method was applied to the PPTA DR1 data set to perform an all-sky search for CWs from individual nearby SMBBHs in circular orbits. Since no GWs were detected, we set upper limits on the intrinsic GW strain amplitude over a range of frequencies. For example, at $10^{-8}$ Hz our analysis has excluded the presence of signals with $h_0$ larger than $1.7\times 10^{-14}$ with $95\%$ confidence. These new limits are a factor of four better than those presented in PPTA10, and a factor of two better than the recent NANOGrav limits reported in \citet{NANOGravCW14}. We also placed upper limits on the coalescence rate of nearby ($z\lesssim 0.1$) SMBBHs, e.g., for very massive binaries ($M_{c}=10^{10} M_{\odot}$) the rate is constrained to be less than $4 \times 10^{-3} {\rm{Mpc}}^{-3} {\rm{Gyr}}^{-1}$ with $95\%$ confidence.

We have also presented all-sky and directional sensitivity curves and find that for the frequency band of $10^{-8}$--$10^{-7}$ Hz (except two narrow bands centred around 1 and 2 $\rm{yr}^{-1}$):\newline
1) With the current data set we are able to detect with $95\%$ probability very massive binary systems ($M_{c}=10^{10} M_{\odot}$) out to a luminosity distance of 400 Mpc regardless of their sky locations and orientations;\newline
2) For the most sensitive sky direction, the current data set is sensitive to average-oriented SMBBHs of chirp masses $10^{9} M_{\odot}$ up to about 100 Mpc.\newline
Furthermore, we show a PPTA sensitivity map in Fig. \ref{fig:SenSkyF2} and find that the PPTA DR1 data set is sensitive to potential SMBBHs with orbital frequencies between 5 and 50 nHz and chirp masses of $\gtrsim 2.3\times 10^{9} M_{\odot}$ in the Coma Cluster, and of $\gtrsim 7\times 10^{8} M_{\odot}$ for both the Fornax Cluster and Virgo Cluster. Directional sensitivity curves and the sensitivity sky map presented here can be used as a guide for future directional/targeted searches. Constraints on specific individual SMBBH candidates will be investigated in a future paper.

Finally, it should be emphasized that our results are based on the assumption that black hole binaries are in circular orbits. Recent models for the SMBBH population including the effects of binary environments on orbital evolution suggest that this assumption is incorrect for GW frequencies $\lesssim 10^{-8}$ Hz \citep{Sesana13CQG,Ravi14GWB}. For future searches, orbital eccentricities and effects of spins will be considered and detection methods optimized for such sources are in development.
\section*{Acknowledgments}
We thank the anonymous referee for very useful comments on the manuscript. The Parkes radio telescope is part of the Australia Telescope National Facility which is funded by the Commonwealth of Australia for operation as a National Facility managed by CSIRO. The work was supported by iVEC through the use of advanced computing resources located at iVEC@UWA. X-JZ thanks S. Chung and Y. Wang for useful discussions and acknowledges the support of an University Postgraduate Award at UWA. X-JZ additionally acknowledges the support for his attendance at the IPTA@Banff meeting in 2014 June from an University student travel award and a PSA travel award at UWA. GH is supported by an Australian Research Council Future Fellowship. LW acknowledges funding support from the Australian Research Council. J-BW is supported by West Light Foundation of CAS XBBS201322 and NSFC project No.11403086. VR is a recipient of a John Stocker Postgraduate Scholarship from the Science and Industry Endowment Fund.

\bibliographystyle{mn2e}
\bibliography{PPTAcw}

\begin{thebibliography}{54}
\expandafter\ifx\csname natexlab\endcsname\relax\def\natexlab#1{#1}\fi

\bibitem[{{Arzoumanian} {et~al}\mbox{.}(2014){Arzoumanian}, {Brazier},
  {Burke-Spolaor}, {Chamberlin}, {Chatterjee}, {Cordes}, {Demorest}, {Deng},
  {Dolch}, {Ellis}, {Ferdman}, {Garver-Daniels}, {Jenet}, {Jones}, {Kaspi},
  {Koop}, {Lam}, {Lazio}, {Lommen}, {Lorimer}, {Luo}, {Lynch}, {Madison},
  {McLaughlin}, {McWilliams}, {Nice}, {Palliyaguru}, {Pennucci}, {Ransom},
  {Sesana}, {Siemens}, {Stairs}, {Stinebring}, {Stovall}, {Swiggum},
  {Vallisneri}, {van Haasteren}, {Wang}, \& {Zhu}}]{NANOGravCW14}
{Arzoumanian} Z. {et~al.}, 2014, arXiv:1404.1267

\bibitem[{{Babak} \& {Sesana}(2012)}]{Babak2012}
{Babak} S., {Sesana} A., 2012, \prd, 85, 044034

\bibitem[{{Burke-Spolaor}(2013)}]{SarahCQG13}
{Burke-Spolaor} S., 2013, Classical and Quantum Gravity, 30, 224013

\bibitem[{{Coles} {et~al}\mbox{.}(2011){Coles}, {Hobbs}, {Champion}, \&
  et~al.}]{Coles2011}
{Coles} W., {Hobbs} G., {Champion} D.~J., et~al., 2011, MNRAS, 418, 561

\bibitem[{{Conselice}(2014)}]{ARAA14}
{Conselice} C.~J., 2014, \araa, 52, 291

\bibitem[{{Corbin} \& {Cornish}(2010)}]{CorbinCornish10}
{Corbin} V., {Cornish} N.~J., 2010, arXiv:1008.1782

\bibitem[{{Demorest} {et~al}\mbox{.}(2013){Demorest}, {Ferdman}, {Gonzalez},
  {Nice}, {Ransom}, {Stairs}, {Arzoumanian}, {Brazier}, {Burke-Spolaor},
  {Chamberlin}, {Cordes}, {Ellis}, {Finn}, {Freire}, {Giampanis}, {Jenet},
  {Kaspi}, {Lazio}, {Lommen}, {McLaughlin}, {Palliyaguru}, {Perrodin},
  {Shannon}, {Siemens}, {Stinebring}, {Swiggum}, \& {Zhu}}]{NANOGrav2012}
{Demorest} P.~B. {et~al.}, 2013, \apj, 762, 94

\bibitem[{{Deng} \& {Finn}(2011)}]{DengXH2011}
{Deng} X., {Finn} L.~S., 2011, \mnras, 414, 50

\bibitem[{{Edwards} {et~al}\mbox{.}(2006){Edwards}, {Hobbs}, \&
  {Manchester}}]{TEMPO2Edwards}
{Edwards} R.~T., {Hobbs} G.~B., {Manchester} R.~N., 2006, \mnras, 372, 1549

\bibitem[{{Ellis}(2013)}]{EllisBayesian}
{Ellis} J.~A., 2013, Classical and Quantum Gravity, 30, 224004

\bibitem[{{Ellis} {et~al}\mbox{.}(2012){Ellis}, {Siemens}, \&
  {Creighton}}]{Ellis2012}
{Ellis} J.~A., {Siemens} X., {Creighton} J.~D.~E., 2012, \apj, 756, 175

\bibitem[{{Foster} \& {Backer}(1990)}]{Foster_Backer90}
{Foster} R.~S., {Backer} D.~C., 1990, \apj, 361, 300

\bibitem[{{Hellings}(1981)}]{Hellings81}
{Hellings} R.~W., 1981, \prd, 23, 832

\bibitem[{{Hellings} \& {Downs}(1983)}]{Hellings_Downs83}
{Hellings} R.~W., {Downs} G.~S., 1983, \apjl, 265, L39

\bibitem[{{Hobbs}(2013)}]{PPTA13CQG}
{Hobbs} G., 2013, Classical and Quantum Gravity, 30, 224007

\bibitem[{{Hobbs} {et~al}\mbox{.}(2010){Hobbs}, {Archibald}, {Arzoumanian},
  {Backer}, {Bailes}, {Bhat}, {Burgay}, {Burke-Spolaor}, {Champion}, {Cognard},
  {Coles}, {Cordes}, {Demorest}, {Desvignes}, {Ferdman}, {Finn}, {Freire},
  {Gonzalez}, {Hessels}, {Hotan}, {Janssen}, {Jenet}, {Jessner}, {Jordan},
  {Kaspi}, {Kramer}, {Kondratiev}, {Lazio}, {Lazaridis}, {Lee}, {Levin},
  {Lommen}, {Lorimer}, {Lynch}, {Lyne}, {Manchester}, {McLaughlin}, {Nice},
  {Oslowski}, {Pilia}, {Possenti}, {Purver}, {Ransom}, {Reynolds}, {Sanidas},
  {Sarkissian}, {Sesana}, {Shannon}, {Siemens}, {Stairs}, {Stappers},
  {Stinebring}, {Theureau}, {van Haasteren}, {van Straten}, {Verbiest},
  {Yardley}, \& {You}}]{IPTA}
{Hobbs} G. {et~al.}, 2010, Classical and Quantum Gravity, 27, 084013

\bibitem[{{Hobbs} {et~al}\mbox{.}(2012){Hobbs}, {Coles}, {Manchester}, {Keith},
  {Shannon}, {Chen}, {Bailes}, {Bhat}, {Burke-Spolaor}, {Champion},
  {Chaudhary}, {Hotan}, {Khoo}, {Kocz}, {Levin}, {Oslowski}, {Preisig}, {Ravi},
  {Reynolds}, {Sarkissian}, {van Straten}, {Verbiest}, {Yardley}, \&
  {You}}]{George_Clock12}
{Hobbs} G. {et~al.}, 2012, \mnras, 427, 2780

\bibitem[{{Hobbs} {et~al}\mbox{.}(2006){Hobbs}, {Edwards}, \&
  {Manchester}}]{TEMPO2}
{Hobbs} G.~B., {Edwards} R.~T., {Manchester} R.~N., 2006, \mnras, 369, 655

\bibitem[{{Iguchi} {et~al}\mbox{.}(2010){Iguchi}, {Okuda}, \&
  {Sudou}}]{3C66B10ApJ}
{Iguchi} S., {Okuda} T., {Sudou} H., 2010, \apjl, 724, L166

\bibitem[{{Jaffe} \& {Backer}(2003)}]{Jaffe_Backer03}
{Jaffe} A.~H., {Backer} D.~C., 2003, \apj, 583, 616

\bibitem[{{Jaranowski} \& {Kr{\'o}lak}(2012)}]{GWDA_LRR12}
{Jaranowski} P., {Kr{\'o}lak} A., 2012, Living Reviews in Relativity, 15, 4

\bibitem[{{Jaranowski} {et~al}\mbox{.}(1998){Jaranowski}, {Kr{\'o}lak}, \&
  {Schutz}}]{Fsts98}
{Jaranowski} P., {Kr{\'o}lak} A., {Schutz} B.~F., 1998, \prd, 58, 063001

\bibitem[{{Jenet} {et~al}\mbox{.}(2006){Jenet}, {Hobbs}, {van Straten}, \&
  et~al.}]{Jenet2006}
{Jenet} F.~A., {Hobbs} G.~B., {van Straten} W., et~al., 2006, ApJ, 653, 1571

\bibitem[{{Jenet} {et~al}\mbox{.}(2004){Jenet}, {Lommen}, {Larson}, \&
  {Wen}}]{JenetWen04}
{Jenet} F.~A., {Lommen} A., {Larson} S.~L., {Wen} L., 2004, \apj, 606, 799

\bibitem[{{Keith} {et~al}\mbox{.}(2013){Keith}, {Coles}, {Shannon}, {Hobbs},
  {Manchester}, {Bailes}, {Bhat}, {Burke-Spolaor}, {Champion}, {Chaudhary},
  {Hotan}, {Khoo}, {Kocz}, {Os{\l}owski}, {Ravi}, {Reynolds}, {Sarkissian},
  {van Straten}, \& {Yardley}}]{MikeDM2013}
{Keith} M.~J. {et~al.}, 2013, \mnras, 429, 2161

\bibitem[{{Kramer} \& {Champion}(2013)}]{EPTA}
{Kramer} M., {Champion} D.~J., 2013, Classical and Quantum Gravity, 30, 224009

\bibitem[{{Lee} {et~al}\mbox{.}(2011){Lee}, {Wex}, {Kramer}, {Stappers},
  {Bassa}, {Janssen}, {Karuppusamy}, \& {Smits}}]{KJLee2011}
{Lee} K.~J., {Wex} N., {Kramer} M., {Stappers} B.~W., {Bassa} C.~G., {Janssen}
  G.~H., {Karuppusamy} R., {Smits} R., 2011, \mnras, 414, 3251

\bibitem[{{Manchester}(2013)}]{IPTAdick13}
{Manchester} R.~N., 2013, Classical and Quantum Gravity, 30, 224010

\bibitem[{{Manchester} {et~al}\mbox{.}(2013){Manchester}, {Hobbs}, {Bailes},
  {Coles}, {van Straten}, {Keith}, {Shannon}, {Bhat}, {Brown}, {Burke-Spolaor},
  {Champion}, {Chaudhary}, {Edwards}, {Hampson}, {Hotan}, {Jameson}, {Jenet},
  {Kesteven}, {Khoo}, {Kocz}, {Maciesiak}, {Oslowski}, {Ravi}, {Reynolds},
  {Sarkissian}, {Verbiest}, {Wen}, {Wilson}, {Yardley}, {Yan}, \&
  {You}}]{PPTA2013}
{Manchester} R.~N. {et~al.}, 2013, \pasa, 30, 17

\bibitem[{{Manchester} {et~al}\mbox{.}(2005){Manchester}, {Hobbs}, {Teoh}, \&
  {Hobbs}}]{ATNF05Pulsar}
{Manchester} R.~N., {Hobbs} G.~B., {Teoh} A., {Hobbs} M., 2005, \aj, 129, 1993

\bibitem[{{McLaughlin}(2013)}]{NANOGrav}
{McLaughlin} M.~A., 2013, Classical and Quantum Gravity, 30, 224008

\bibitem[{{Petiteau} {et~al}\mbox{.}(2013){Petiteau}, {Babak}, {Sesana}, \& {de
  Ara{\'u}jo}}]{Petiteau13}
{Petiteau} A., {Babak} S., {Sesana} A., {de Ara{\'u}jo} M., 2013, \prd, 87,
  064036

\bibitem[{{Ravi} {et~al}\mbox{.}(2012){Ravi}, {Wyithe}, {Hobbs}, \&
  et~al.}]{Ravi2012}
{Ravi} V., {Wyithe} J.~S.~B., {Hobbs} G., et~al., 2012, ApJ, 761, 84

\bibitem[{{Ravi} {et~al}\mbox{.}(2014{\natexlab{a}}){Ravi}, {Wyithe},
  {Shannon}, \& {Hobbs}}]{Ravi14Single}
{Ravi} V., {Wyithe} J.~S.~B., {Shannon} R.~M., {Hobbs} G., 2014{\natexlab{a}},
  arXiv:1406.5297

\bibitem[{{Ravi} {et~al}\mbox{.}(2014{\natexlab{b}}){Ravi}, {Wyithe},
  {Shannon}, {Hobbs}, \& {Manchester}}]{Ravi14GWB}
{Ravi} V., {Wyithe} J.~S.~B., {Shannon} R.~M., {Hobbs} G., {Manchester} R.~N.,
  2014{\natexlab{b}}, \mnras, 442, 56

\bibitem[{{Rosado} \& {Sesana}(2014)}]{Rosado14}
{Rosado} P.~A., {Sesana} A., 2014, \mnras, 439, 3986

\bibitem[{{Sesana}(2013{\natexlab{a}})}]{Sesana13CQG}
{Sesana} A., 2013{\natexlab{a}}, Classical and Quantum Gravity, 30, 224014

\bibitem[{{Sesana}(2013{\natexlab{b}})}]{Sesana13GWB}
{Sesana} A., 2013{\natexlab{b}}, \mnras, 433, L1

\bibitem[{{Sesana} \& {Vecchio}(2010)}]{Sesana2010}
{Sesana} A., {Vecchio} A., 2010, \prd, 81, 104008

\bibitem[{{Sesana} {et~al}\mbox{.}(2009){Sesana}, {Vecchio}, \&
  {Volonteri}}]{Sesana2009}
{Sesana} A., {Vecchio} A., {Volonteri} M., 2009, MNRAS, 394, 2255

\bibitem[{{Shannon} {et~al}\mbox{.}(2013){Shannon}, {Ravi}, {Coles}, {Hobbs},
  {Keith}, {Manchester}, {Wyithe}, {Bailes}, {Bhat}, {Burke-Spolaor}, {Khoo},
  {Levin}, {Oslowski}, {Sarkissian}, {van Straten}, {Verbiest}, \&
  {Wang}}]{PPTA2013Sci}
{Shannon} R.~M. {et~al.}, 2013, Science, 342, 334

\bibitem[{{Simon} {et~al}\mbox{.}(2014){Simon}, {Polin}, {Lommen}, {Stappers},
  {Finn}, {Jenet}, \& {Christy}}]{SimonGWhot14}
{Simon} J., {Polin} A., {Lommen} A., {Stappers} B., {Finn} L.~S., {Jenet}
  F.~A., {Christy} B., 2014, \apj, 784, 60

\bibitem[{{Sundelius} {et~al}\mbox{.}(1997){Sundelius}, {Wahde}, {Lehto}, \&
  {Valtonen}}]{OJ287ApJ97}
{Sundelius} B., {Wahde} M., {Lehto} H.~J., {Valtonen} M.~J., 1997, \apj, 484,
  180

\bibitem[{{Taylor} {et~al}\mbox{.}(2014){Taylor}, {Ellis}, \&
  {Gair}}]{Taylor14}
{Taylor} S., {Ellis} J., {Gair} J., 2014, arXiv:1406.5224

\bibitem[{Thorne(1987)}]{Thorne87}
Thorne K.~S., 1987, Gravitational Radiation, in 300 years of Gravitation,
  edited by S. Hawking and W. Israel, pages 330-458. Cambridge University
  Press, Cambridge, UK

\bibitem[{{Valtonen} {et~al}\mbox{.}(2010){Valtonen}, {Mikkola}, {Lehto},
  {Hyv{\"o}nen}, {Nilsson}, {Merritt}, {Gopakumar}, {Rampadarath}, {Hudec},
  {Basta}, \& {Saunders}}]{OJ287mass}
{Valtonen} M.~J. {et~al.}, 2010, Celestial Mechanics and Dynamical Astronomy,
  106, 235

\bibitem[{{van Haasteren} {et~al}\mbox{.}(2011){van Haasteren}, {Levin},
  {Janssen}, \& et~al.}]{EPTAlimit}
{van Haasteren} R., {Levin} Y., {Janssen} G.~H., et~al., 2011, MNRAS, 414, 3117

\bibitem[{{Verbiest} {et~al}\mbox{.}(2009){Verbiest}, {Bailes}, {Coles},
  {Hobbs}, {van Straten}, {Champion}, {Jenet}, {Manchester}, {Bhat},
  {Sarkissian}, {Yardley}, {Burke-Spolaor}, {Hotan}, \& {You}}]{Verbiest09}
{Verbiest} J.~P.~W. {et~al.}, 2009, \mnras, 400, 951

\bibitem[{{Wang} {et~al}\mbox{.}(2014{\natexlab{a}}){Wang}, {Hobbs}, {Coles},
  \& et~al.}]{GWM_PPTA}
{Wang} J.~B., {Hobbs} G.~B., {Coles} W., et~al., 2014{\natexlab{a}}, MNRAS,
  submitted

\bibitem[{{Wang} {et~al}\mbox{.}(2014{\natexlab{b}}){Wang}, {Mohanty}, \&
  {Jenet}}]{YanWang14}
{Wang} Y., {Mohanty} S.~D., {Jenet} F.~A., 2014{\natexlab{b}}, arXiv:1406.5496

\bibitem[{{Wen} {et~al}\mbox{.}(2011){Wen}, {Jenet}, {Yardley}, {Hobbs}, \&
  {Manchester}}]{ZLWen11LimRate}
{Wen} Z.~L., {Jenet} F.~A., {Yardley} D., {Hobbs} G.~B., {Manchester} R.~N.,
  2011, \apj, 730, 29

\bibitem[{{Wyithe} \& {Loeb}(2003)}]{Wyithe_Loeb03}
{Wyithe} J.~S.~B., {Loeb} A., 2003, \apj, 590, 691

\bibitem[{{Yardley} {et~al}\mbox{.}(2011){Yardley}, {Coles}, {Hobbs},
  {Verbiest}, {Manchester}, {van Straten}, {Jenet}, {Bailes}, {Bhat},
  {Burke-Spolaor}, {Champion}, {Hotan}, {Oslowski}, {Reynolds}, \&
  {Sarkissian}}]{YardleySGWB}
{Yardley} D.~R.~B. {et~al.}, 2011, \mnras, 414, 1777

\bibitem[{{Yardley} {et~al}\mbox{.}(2010){Yardley}, {Hobbs}, {Jenet}, \&
  et~al.}]{Yardley2010}
{Yardley} D.~R.~B., {Hobbs} G.~B., {Jenet} F.~A., et~al., 2010, MNRAS, 407, 669

\end{thebibliography}
\label{lastpage}
\end{document}